\documentclass[a4paper,11pt]{article}
\usepackage{geometry,tikz}
\usepackage{a4wide}
\usepackage{graphicx}
\usepackage{epsf}
\usepackage{amsmath}
\usepackage{amssymb}
\usepackage[normalem]{ulem}

\usepackage{cite}\newcommand\orcidtoniato{{\href{https://orcid.org/0000-0003-0969-9359}{\orcidicon}}}
\usepackage{cite}\newcommand\orcidroldao{{\href{https://orcid.org/0000-0003-3978-532X}{\orcidicon}}}
\usepackage{cite}\newcommand\orciddudal{{\href{http://orcid.org/0000-0003-1326-6011}{\orcidicon}}}
\usepackage{cite}\newcommand\orcidhudson{{\href{https://orcid.org/0000-0002-3519-9553}{\orcidicon}}}

\newcommand{\orcidicon}{%
	\begin{tikzpicture}
	\draw[lime, fill=lime] (0,0)
		circle [radius=0.16]
		node[white] {{\fontfamily{qag}\selectfont \tiny ID}};
	\draw[white, fill=white] (-0.0625,0.095)
		circle [radius=0.007];
	\end{tikzpicture}	\hspace{-2mm}
}

\usepackage{multirow}
\usepackage{subfigure}
\usepackage{appendix}
\usepackage{tikz}
\usepackage{amsfonts,amsthm,euscript,braket,xcolor}
\usepackage{breqn}
\newcommand{\be}{\begin{equation}}
	\newcommand{\ee}{\end{equation}}

\newcommand{\Rmnum}[1]{\expandafter\@slowromancap\romannumeral #1@}
\newcommand{\bea}{\begin{eqnarray}}
	\newcommand{\eea}{\end{eqnarray}}

\newcommand{\p}{\partial}
\newcommand\scalemath[2]{\scalebox{#1}{\mbox{\ensuremath{\displaystyle #2}}}}
\numberwithin{equation}{section}

\newcommand\Co[1]\null

\usepackage{chngcntr,ulem}
\counterwithin{table}{section}
\counterwithin{figure}{section}

\usepackage{ulem}
\usepackage[colorlinks=true, linkcolor=blue, citecolor=blue, urlcolor=blue]{hyperref}
\begin{document}
	%%%%%%%%%%%%%%%%%%%%%%%%%%%%%%%
	
	%%%%%%%%%%%%%%%%%%%%%%%%%%%%
	
	\title{\bf A dynamical Einstein--Proca-dilaton holographic model and exotic hybrid meson resonances}
\author{\textbf{Bruno Toniato\orcidtoniato\!\!$^{a}$}\thanks{bruno.toniato@ufabc.edu.br (corresponding author)},
\textbf{David Dudal\orciddudal\!\!$^{b}$}\thanks{david.dudal@kuleuven.be},
\textbf{Hudson Almeida$^{a,b}$}\orcidhudson\thanks{hudson.vinicius@ufabc.edu.br}, \\
\textbf{Roldao da Rocha\orcidroldao\!\!$^{c}$}
\thanks{roldao.rocha@ufabc.edu.br} \\
\\
\textit{\small $^a$ Federal University of ABC, Center of Physics,}\\
\textit{\small Santo Andr\'e, S\~ao Paulo, 09210-170, Brazil}\\
\textit{\small $^b$ KU Leuven Campus Kortrijk--Kulak, Department of Physics, Etienne Sabbelaan 53 bus 7657,}\\
\textit{\small 8500 Kortrijk, Belgium}\\
\textit{\small $^c$ Federal University of ABC, Center of Mathematics,}\\
\textit{\small Santo Andr\'e, S\~ao Paulo, 09210-170, Brazil}
}

\date{}
\maketitle
\begin{abstract}
{\color{black}{A fully dynamical Einstein--Proca-dilaton (EPD) holographic model is developed for the exotic $J^{PC}=1^{-+}$ hybrid meson, incorporating the backreaction of the massive vector field dual to the hybrid current in the bulk. Solving the coupled EPD system yields a self-consistent gravitational background and the mass spectrum of hybrid resonances. While the soft-wall limit reproduces linear Regge behavior, the inclusion of backreaction yields quantitative shifts that improve agreement with experimental candidates. The finite-temperature regime is approached by studying black hole thermodynamics and spectral functions across the confinement–deconfinement transition, fixing the Hawking--Page temperature to $T_c \sim 150\,\mathrm{MeV}$. Unlike non-backreacted AdS/QCD models, the dynamical background predicts a strong suppression of hybrid states already near the onset of deconfinement. These results demonstrate that gravitational backreaction has measurable consequences for both vacuum spectroscopy and in-medium properties of exotic hybrid mesons, emphasizing the relevance of a dynamical self-consistent approach in AdS/QCD.}}
\end{abstract}

\section{Introduction}
\label{sec1}

Quantum chromodynamics (QCD) describes the strong interactions among quarks and gluons. Dealing with QCD in the low-energy regime poses theoretical challenges regarding the strongly coupled behavior of quarks and gluons. The AdS/CFT correspondence has offered a profound reformulation of how to deal with certain strongly coupled quantum field theories \cite{Maldacena:1997re}. It conjectures an exact duality between type IIB superstring theory formulated on a ten-dimensional curved spacetime and a conformal gauge theory defined on its four-dimensional boundary. In its canonical realization, type IIB string theory in the near-horizon geometry underlying a stack of $N_c$ coincident $D_3$-branes, whose low-energy limit is described by the AdS$_5 \times S^5$ background, is holographically dual to $\mathcal{N}=4$ supersymmetric $SU(N_c)$ Yang--Mills theory in four dimensions. Refs.~\cite{Witten:1998qj,Gubser:1998bc,Aharony:1999ti} further clarified how generating functionals of boundary correlation functions are obtained from the on-shell gravitational action. These works established the operational framework that enables AdS/CFT to be computationally powerful. In the large-$N_c$ and large ’t Hooft coupling limit, the strongly coupled gauge theory is mapped to classical supergravity in asymptotically AdS spacetime.

Although $\mathcal{N}=4$ supersymmetric Yang--Mills theory is conformal and does not exhibit confinement, holography-inspired phenomenological constructions aim to capture essential features of QCD. AdS/QCD relates weakly coupled gravity in AdS$_5$ to the nonperturbative regime of QCD by introducing infrared (IR) deformations that emulate confinement and chiral symmetry breaking. Early bottom-up implementations include the hard-wall model \cite{Erlich:2005qh} and the soft-wall construction \cite{Karch:2006pv}, which successfully reproduce hadronic spectra and chiral symmetry breaking \cite{Chelabi:2015gpc}. Extensions addressing meson structure and decay constants were explored in Refs.~\cite{Colangelo:2008us,Vega:2018dgk,Branz:2010ub,MartinContreras:2021bis,Ballon-Bayona:2014oma,Junqueira:2025evt}. Light-front holographic approaches, connecting AdS dynamics to the semiclassical description of hadrons, were developed in Ref.~\cite{dePaula:2009za,Afonin:2022hbb,Brodsky:2014yha}.

Top-down constructions derived directly from string theory, most notably the Sakai--Sugimoto model introduced in Refs.~\cite{Sakai2005Apr}, provide a geometric realization of chiral symmetry breaking in a confining background \cite{Boschi-Filho:2002xih,MartinContreras:2023eft}. Relevant developments and applications to strongly interacting matter can be found, for instance, in Refs.~\cite{MartinContreras:2022lxl,Rougemont:2023gfz,Sorensen:2023zkk}. These approaches report how AdS/QCD can be employed to address nonperturbative QCD dynamics, encoding confinement, mass generation, and hadronic structure in the properties of gravitational backgrounds in AdS space.
In the so-called bottom-up approach, the bulk dynamics are constrained by phenomenological inputs to reproduce desirable features of QCD, such as chiral symmetry breaking and confinement \cite{Gursoy:2007cb,Alanen:2009xs}. Further developments of bottom-up holographic QCD can be found in Refs.~\cite{He:2013qq,Alho:2013hsa,Bartz:2018nzn,daRocha:2021xwq}. More recent analyses appear in Refs.~\cite{Ballon-Bayona:2023zal,Arefeva:2024vom}. Additional modern perspectives are presented in Refs.~\cite{Erdmenger:2020flu,Karapetyan:2023sfo,Almeida:2026ozb}.
Within this context, the soft-wall model introduces a background dilaton field that softly breaks conformal invariance, enabling a realistic description of confinement and chiral symmetry breaking \cite{Fujita:2009wc}. Foundational aspects of this mechanism were discussed in Ref.~\cite{Csaki}, while dynamical implementations appear in Ref.~\cite{Gherghetta:2009ac,Bartz:2024dgd,Ishii:2019gta}.
AdS/QCD models have been extensively applied to the study of the hadronic spectrum \cite{Braga:2017oqw,Braga:2023qee,Bernardini:2016hvx}. Investigations of mesonic states and decay properties were carried out in Refs.~\cite{Ferreira:2019inu,Ferreira:2020iry}. Further studies addressing spectroscopy and related observables appear in Refs.~\cite{Colangelo:2018mrt,Bernardini:2018uuy,daRocha:2021ntm,daRocha:2024sjn,Deng:2025azp}.

One of the persistent difficulties in hadron spectroscopy concerns the proper treatment of gluonic degrees of freedom. This issue becomes particularly relevant in hadrons that contain both valence quarks and explicit gluonic excitations, giving rise to so-called hybrid states. The holographic description of such states requires the identification of appropriate bulk fields dual to gluonic operators, as well as a consistent implementation of their mixing with quark bilinears, an active direction in the application of AdS/QCD to nonperturbative strong-interaction physics \cite{MartinContreras:2020cyg}. 
Exotic mesonic states have been postulated since the advent of QCD \cite{JAFFE1976201}. Such states cannot be described by the usual quark model content, where mesons are composed of a quark-antiquark pair $(q\bar{q})$. The conventional mesons have parity $P=(-1)^{L+1}$ and charge conjugation $C=(-1)^{L+S}$, with $L$ and $S$ denoting the relative angular momentum and the total spin of the pair $(q\bar{q})$. Thus, $J^{PC}$ combinations that cannot be reached with any value of $L$ and $S$, such as $0^{+-},1^{-+},$ and $2^{+-}$, are classified as exotic ones  \cite{Jaffe1}. Unconventional $J^{PC}$ combinations can be implemented by adding a gluonic excitation $G$ to the pair $q\bar{q}$, leading to a mesonic hybrid state $qG\bar{q}$.  Hybrid mesons have been analyzed in the context of the MIT bag model in Refs.~\cite{Chanowitz:1982qj,BARNES1983241} and the flux-tube model in Refs.~\cite{PhysRevD.31.2910,PhysRevD.52.5242}. An effective field theory has also been developed to describe hybrid mesons   \cite{Berwein:2015vca}. In addition, a  comprehensive investigation of the lightest exotic $1^{-+}$ hybrid meson has emerged over the past decade. Early lattice QCD investigations established the structure of hybrid mesons  \cite{Dudek:2011bn,Burns:2006wz,Woss:2020ayi,Chen:2022isv}, while QCD sum-rule analyses provided refined mass predictions \cite{Huang:2014hya,Esmer:2025xss}. The pole position and further phenomenological aspects of hybrid mesons were developed in Refs. \cite{JPAC:2018zyd,Swanson:2023zlm}. 
From the experimental side, no exotic hybrid candidate has yet been observed in the heavy-flavor sector, despite theoretical developments \cite{Miyamoto:2019oin}. On the other hand, in the light-flavor sector, the $J^{PC}=1^{-+}$ hybrid mesons $\pi_1(1400)$, $\pi_1(1600)$, and $\pi_1(2015)$ have been reported from experimental collaborations \cite{ParticleDataGroup:2024cfk,E852:1999xev,PhysRevLett.81.5760,PhysRevLett.104.241803,PhysRevLett.94.032002}.

The $J^{PC}=1^{-+}$ hybrid meson has been previously studied in phenomenological AdS/QCD models, both in the hard-wall \cite{Kim:2008qh} and in the soft-wall \cite{Bellantuono:2014lra,Rinaldi:2024fgx} approaches. In the soft-wall framework, the $1^{-+}$ hybrid state dissociates at lower temperatures compared to conventional light-flavor mesons, indicating greater instability. Here, we investigate hybrid mesons by a self-consistent dynamical approach based on the Einstein--Maxwell-dilaton (EMD) gravity framework, which has successfully described the mass spectrum and melting behavior of exotic mesons at finite temperature and density \cite{Toniato:2025gts}. Ref.~\cite{Rougemont:2015ona} showed that the EMD model can match lattice QCD data for the quark-gluon plasma (QGP) equation of state near the crossover transition, up to baryon chemical potentials of 400~MeV. The EMD system has also been widely employed to study QCD phenomena \cite{Noronha:2010hb,He2013Apr,Ballon-Bayona:2020xls,Gubser:2008yx,Bohra2020Feb,Bohra2021Apr,Dudal:2021jav,Cai:2012xh,Yang:2015aia,Jena2022Apr,Shukla:2023pbp,Shukla:2024qlf}, and Ref.~\cite{Dudal:2017max} demonstrated that it supports a first-order phase transition between small and large black holes, dual to confinement and deconfinement, allowing the study of confined physics via the small black hole configuration.

In this work, we develop a holographic description of $1^{-+}$ hybrid mesons within a fully backreacted Einstein--Proca-dilaton (EPD) framework and study how gravitational backreaction affects their vacuum and in-medium properties. We construct the five-dimensional EPD action, solve the coupled equations of motion through a potential-reconstruction method, and compute the zero-temperature mass spectrum by reducing the Proca dynamics to a Schrödinger-like problem. The analysis is then extended to finite-temperature self-consistent AdS/QCD by examining black hole thermodynamics, fixing model parameters through the Hawking--Page transition, and evaluating spectral functions in both non-backreacted and fully backreacted geometries, thereby assessing the role of dynamical consistency in the thermal behavior and melting of hybrid mesons.

This paper is organized as follows: Section~\ref{sec2} constructs the holographic framework based on the EPD action, where the five-dimensional gravitational background is coupled to a massive 1-form field dual to the QCD hybrid current and a dilaton. From the action, we derive the coupled Einstein, Proca, and Klein--Gordon equations and implement a potential-reconstruction strategy using suitable Ansätze for the warp factor and kinetic function, ensuring a self-consistent backreacted geometry. Section~\ref{sec3} is devoted to studying the $1^{-+}$ hybrid meson spectrum at zero temperature. Transforming the Proca equation into a Schrödinger-like form, we first recover the Regge behavior in the soft-wall limit and then solve the fully backreacted EPD equations numerically. A comparison shows that backreaction leads to quantitative shifts in the spectrum, improving agreement with experimental data while preserving the qualitative structure of potentials and wavefunctions. Section~\ref{sec4} discusses the implications of these results for the holographic description of hybrid mesons, emphasizing the role of backreaction in shaping the spectrum and enhancing phenomenological consistency. We also outline possible extensions, including finite temperature and density, and applications to other exotic hadrons. Section~\ref{secSource} addresses the interpretation of a finite boundary source for the Proca field. We clarify that, unlike the Maxwell case, such a source should not be identified with a chemical potential, but rather with an external deformation in the exotic hybrid channel. We discuss the corresponding near-boundary expansion, the role of the source and response coefficients, and the reasons why the fully backreacted sourced problem is both conceptually subtle and numerically stiff. Finally, Section~\ref{sec5} concludes by summarizing the main findings and suggesting future directions, such as exploring additional hybrid states, refining the holographic potential, and connecting to lattice QCD results.

\section{The Einstein--Proca-dilaton action}
\label{sec2}
In this work, we start with the EPD action, systematically building a self-consistent holographic framework that can accurately capture the properties of hybrid $1^{-+}$ mesons for their dual field. In QCD the hybrid state $1^{-+}$ can be described by the local gauge invariant operator $J^a_\mu = \bar{q}T^aG_{\mu \nu}\gamma^\nu q$, with $G_{\mu\nu}$ denoting the gluon field strength and $T^a$ flavor matrices such that Tr$[T^aT^b] = \delta^{ab}/2$. This QCD operator has a dual field which is a 1-form, $H_M = H^a_MT^a$ \cite{Bellantuono:2014lra}. In five-dimensional spacetime, the EPD action is written explicitly as follows
\begin{equation}\label{epd_action}
    \scalemath{.93}{S_{\text{EPD}} = \frac{1}{16 \pi G_5} \int d^5x \, \sqrt{-g}
    \left[
        R - \frac{f(\phi)}{4} F_{MN}F^{MN} - \frac{f(\phi)}{2}m_5^2 H_M H^M
        - \frac{1}{2} \partial_M \phi \partial^M \phi
        - V(\phi)
    \right]},
\end{equation}
where $F_{MN} = \partial_M H_N - \partial_N H_M$ is the field strength, $f(\phi)$ is a kinetic function that couples the dilaton to the Proca field $H_M$, which has mass $m_5$, $V(\phi)$ is the dilaton field potential, and $G_5$ is Newton's gravitational constant in five dimensions. Varying the action (\ref{epd_action}) yields  the following equations of motion
\begin{eqnarray}\label{einstein_eq}
    R_{MN} - \frac{1}{2} g_{MN} R &=& T_{MN}\,,\\
\label{proca_eq}
    \nabla_M \left[ f(\phi) F^{MN} \right] - m_5^2 f(\phi) H^N &=& 0\,,\\\label{KG_eq}
    \partial_M \left[ \sqrt{-g} \partial^M \phi \right]
    - \sqrt{-g}
    \left[
        \frac{\partial V}{\partial \phi}
        + \frac{1}{2}\left( m_5^2H^2 +  \frac{F^2}{2} \right) \frac{\partial f}{\partial \phi} 
    \right] &=& 0\,,
\end{eqnarray}
where $F^2=F_{MN}F^{MN}$. Eqs.~(\ref{einstein_eq}) are Einstein's field equations, whereas Eq.~(\ref{proca_eq}) represents the Proca equation, and Eq.~(\ref{KG_eq}) is the Klein--Gordon equation for the dilaton field. The energy-momentum tensor takes the form
\begin{eqnarray}\label{EM_tensor}
  T_{MN} &=& \frac{1}{2} \left[ \partial_M \phi \partial_N \phi - \frac{1}{2}  \left(\partial \phi \right)^2 g_{MN} -  V(\phi) g_{MN} \right]+\frac{f(\phi)}{2} \left[ F_{MP} F^{\;P}_{N}-\frac{1}{4}  F^2  g_{MN}\right]\nonumber\\&&   + \frac{m_5^2 f(\phi)}{2} \left[H_M H_N - \frac{1}{2} H^2 g_{MN}\right]\,.
\end{eqnarray}

In order to construct a self-consistent holographic model, Eqs.~\eqref{einstein_eq} -- \eqref{KG_eq}, which are coupled, must be solved simultaneously. This may be implemented via the potential reconstruction approach \cite{Mahapatra:2020wym,Mahapatra:2018gig,Priyadarshinee:2023cmi,Priyadarshinee:2021rch,Daripa:2024ksg,Toniato:2025gts,Jena:2024cqs}. Accordingly, we adopt the following Ans\"atze for the metric, massive spin-$1$, and dilaton fields,
\begin{equation}\label{ansatz_metric}
    ds^2 = \frac{L^2 e^{2A(z)}}{z^2}
            \left(-g(z) \, dt^2 + \frac{dz^2}{g(z)} + dx_i dx^i \right)\,,
\end{equation}
\begin{equation}
    H_M dx^M = H_t(z) \,dt\,, \quad \phi=\phi(z) \,.
\end{equation}
where $L$ is the AdS radius, $A(z)$ is the warp factor and $g(z)$ is the blackening function. It is assumed that the nonzero components of the spin-1 field $H_t$ and the dilaton field $\phi$ depend only on the holographic coordinate $z \in [0, z_h]$. Here, $z_h$ is the horizon radius defined by $g(z_h)=0$. The strongly coupled gauge theory lives at the asymptotic boundary of spacetime, which is defined by the $z\to0$ limit. Plugging the Ans\"atze (\ref{ansatz_metric}) into Eqs.~(\ref{einstein_eq}) --  (\ref{KG_eq})  yields the following expressions:
\begin{align}
    \label{EPD_eqs}
    0 &= \phi'' + \left(\frac{g'}{g} + 3A' - \frac{3}{z}\right) \phi'
    - \frac{L^2 e^{2A}}{g z^2} \frac{\partial V}{\partial \phi}
    + \frac{z^2 e^{-2A}}{2g L^2} \frac{\partial f}{\partial \phi} (H_t')^2
    + \frac{m_5^2}{2 g^2} \frac{\partial f}{\partial \phi} H_t^2 \,, \\ 
    0 &= H_t'' + \left(\frac{f' \phi'}{f} + A' - \frac{1}{z}\right) H_t'
    - \frac{m_5^2 L^2 e^{2A}}{g z^2} H_t \,, \label{proc} \\ 
    0 &= g'' + g' \left(3A' - \frac{3}{z}\right)
    - \frac{z^2 e^{-2A} f}{L^2} (H_t')^2
    - \frac{m_5^2 f}{g} H_t^2 \,, \label{metriceq} \\  
    0 &= A'' + \frac{g''}{6g}
    + \left(\frac{3 g'}{2g} - \frac{6}{z}\right) A'
    - \frac{1}{z} \left(\frac{3 g'}{2g} - \frac{4}{z}\right)
    + 3(A')^2 + \frac{L^2 e^{2A}}{3 g z^2} V
    - \frac{m_5^2 f}{6 g^2} H_t^2 \,, \label{2.10} \\
    0 &= A'' - A' \left(A' - \frac{2}{z}\right)
    + \frac{(\phi')^2}{6} + \frac{m_5^2 f}{6 g^2} H_t^2 \,.
\end{align}

With five equations for four unknown functions, we take Eq.~(\ref{EPD_eqs}) as a constraint. As shown in Ref. \cite{Dudal:2017max}, considering $A(z)$ and $f(z)$ as free functions enables one to obtain all remaining unknowns in closed form. In particular, imposing the following boundary conditions
\begin{equation}
    \lim_{z \to 0} g(z) = 1 \,, \qquad\qquad  \lim_{z \to z_h} g(z) = 0\,,  \qquad  \qquad \lim_{z \to z_h} H_t(z) = 0\,.
\end{equation}
it is possible to solve the coupled system of equations after the specification of $f(z)$ and $A(z)$.

%%%%%%%%%%%%%%%%%%%%%%%%%%%%%%%%%%%%%%%%%%%%
\section{The mass spectra in the EPD model}
\label{sec3}
In this section, the mass spectroscopy of $1^{-+}$ state will be scrutinized at zero temperature.

\subsection{Choice of the free $A(z)$ and $f(z)$ functions}\label{modelsec}
We will determine the function $f(z)$ first by studying the meson mass spectra at zero temperature. As has been observed in earlier works \cite{Mamani:2022qnf, Dudal:2017max}, an especially suitable choice is
\begin{equation}
    f(z)=e^{-\left[A(z)+B(z)\right]}\,.
\end{equation}

Here, $B(z)$ is fixed to reproduce the mesonic mass spectra. In the original soft-wall model \cite{Herzog:2006ra}, $B(z)$ is identified with the dilaton profile $\Phi(z)$, namely $B(z)\equiv\Phi(z)=c^2 z^2$, where $c$ is a constant that sets the slope of the Regge trajectory.\footnote{Note that the soft-wall dilaton $\Phi(z)$ should not be confused with the dilaton field $\phi(z)$ appearing in our holographic model in Eq.~(\ref{epd_action}).} We adopt this form to match the meson spectra, choosing $c$ in particular to facilitate comparison with the $\pi_1(1400)$ mass in the fully backreacted geometry. We, therefore, from now on, take $c=0.55\,\text{GeV}$ unless stated otherwise.

The remaining input, $A(z)$, may be taken as a simple quadratic profile, whose implications were examined in \cite{Dudal:2017max}. Within this setup, such a choice for $A(z)$ is able to realize confined-deconfined phases in the dual boundary theory, thereby providing a QCD-like description. We therefore employ
\begin{equation}
    A(z)=-\frac{a}{8}\,z^2\,.
\end{equation}
The parameter $a$ will be fixed in Sec.~\ref{sec42} by requiring that the Hawking--Page geometric transition takes place at approximately $T\simeq 150\,\text{MeV}$. This transition temperature is motivated by the QCD lattice estimate of Ref. \cite{Steinbrecher:2018phh}. This implies, as seen in Sec.~\ref{sec42}, that $a = 0.395 \, \text{GeV}^2$, which will be used from now on, unless stated otherwise. 

\subsection{Equation of motion for the hybrid mesons}
Once the two free functions have been specified, we can proceed to compute the meson mass spectra at zero temperature. Starting from the Proca equation~(\ref{proca_eq}) derived from the action~(\ref{epd_action}), we can write
\begin{equation} \label{EOM_1}
%\nabla_M \left[ f(\phi) F^{MN} \right] - m_5^2 f(\phi) H^N = 0 \implies 
\partial_M \left( \sqrt{-g} f(\phi) F^{MN} \right) -  m_5 ^2 f(\phi)\sqrt{-g} H^N  = 0 \,,
\end{equation}
In the zero-temperature limit, the limit $g(z) \rightarrow 1$ holds. Therefore, the spacetime squared line element becomes
\begin{align}\label{metric_T0}
    ds^2 &= \frac{L^2 e^{2A(z)}}{z^2}\left(-\, dt^2 + dz^2 + dx_i dx^i\right)\,.
\end{align}

In 5D, the massive spin-1 field has an additional component compared with the usual 4D case. The extra component of the bulk field corresponds to spin-0 hybrid mesons at the boundary. The action \eqref{epd_action} does not admit gauge invariance, and therefore, this extra degree of freedom cannot be gauged away. Since there is no additional dual QCD operator to which it can be mapped, we will only consider the subclass of classical solutions with $H_z=0$ to effectively remove this obsolete spin-0 component. See also \cite{Zhao:2026bsc} for more about this. With this condition, the $z$-component of Eq.~(\ref{EOM_1}) yields the transverse condition $\p_\mu H^\mu=0$. Thus, the Proca equation becomes
\begin{equation} \label{EOM_2}
\frac{z e^{-A(z)}}{L f(z)} \partial_z \left(\frac{L e^{A(z)} f(z)}{z} \partial_z H_\mu \right) + \Box H_\mu  - \frac{m_5^2L^2e^{2A(z)}}{z^2} H_\mu = 0 \,,
\end{equation}
which can be led into a Schr\"odinger-like form by first implementing a Fourier transform
\begin{equation}
\label{fourier}
H_\mu(z, x^\nu) = \int \frac{d^4k}{(2\pi)^4} e^{i k_\alpha x^\alpha} H_\mu(z, k)\,,
\end{equation}
and performing the following change of variables
\begin{align} \label{B_def}
    H^\mu &= e^{-\xi} \psi^\mu\,, \\
    \xi &= \frac{1}{2} \log\left[\frac{e^{A(z)} f(z)}{z}\right]\,.
\end{align}

This turns Eq.~(\ref{EOM_2}) into
\begin{equation} \label{Mass_Spectra_T0}
-\partial_z^2 \psi_n(z) + V(z) \psi(z) = m_n^2 \psi(z)\,,
\end{equation}
for the potential 
\begin{equation} \label{Potential_T0}
V(z) = \left(\partial_z \left[\frac{1}{2} \log\left(\frac{e^{A(z)} f(z)}{z}\right)\right]\right)^2
+ \partial_z^2 \left[\frac{1}{2} \log\left(\frac{e^{A(z)} f(z)}{z}\right)\right] +\frac{m_5^2L^2e^{2A(z)}}{z^2} \,,
\end{equation}
where we relate the eigenmasses via $k^2=-m_n^{2}=-\omega^2+p^2$. Moreover, the Proca field, which is dual to a local QCD operator, carries a bulk mass fixed by the AdS/CFT dictionary as $m_5^2L^2=(\Delta-p)(\Delta+p-4)$ for a $p$-form field dual to an operator of scaling dimension $\Delta$. In the present case, the relevant local QCD operator describing a hybrid $1^{-+}$ meson is the current $J_\mu^{a}=\bar{q}\,T^{a}G_{\mu\nu}\gamma^{\nu}q$. It then follows that, for a $1$-form Proca field with $\Delta=5$, one obtains $m_5^2L^2=(5-1)(5+1-4)=8$.

Eq.~(\ref{Mass_Spectra_T0}) is equivalent to the result found in Ref.  \cite{Bellantuono:2014lra} if one takes the warping of AdS$_5$ geometry to zero, $A(z) \to 0$. This implies the original soft-wall model geometry (pure AdS$_5$). Considering first this case as a warm-up exercise, Eq. (\ref{Mass_Spectra_T0}) takes the form 
\begin{gather} 
-\partial_z^2 \psi_n(z) + \left[\left(\partial_z \left[\frac{1}{2} \log\left(\frac{e^{-c^2z^2}}{z}\right)\right]\right)^2
+ \partial_z^2 \left[\frac{1}{2} \log\left(\frac{e^{-c^2z^2}}{z}\right)\right] +\frac{8}{z^2} \right] \psi_n(z) = m_n^2 \psi(z)\,,
\end{gather}
which can be recast as 
\begin{gather}
 -\partial_z^2 \psi_n(z) + \left[\frac{35}{4z^2} + c^4z^2\right] \psi_n(z)  = m_n^2 \psi_n(z) \,.\label{harmonic}
\end{gather}
We identify Eq.~(\ref{harmonic}) as the three-dimensional radial harmonic oscillator equation, with effective potential $V(r) = \omega^2r^2 + l(l+1)/r^2$, so its normalized solution reads
\begin{equation}
    \psi_{nl}(z) = \mathcal{N}_n \left(cz \right)^{l+1} e^{-c^2z^2/2} \, L_n^{l+1/2}(c^2z^2)\,,
    \end{equation}
    with corresponding mass spectrum 
    \begin{equation}
    m_{nl}^2 = 2 c^2 \left(2n+l+\frac{3}{2} \right)\,,
\end{equation}
where $l(l+1) = 35/4$, implying that $l=5/2$, where  $L_n^{l+1/2}$ denotes the generalized Laguerre polynomial. We therefore find 
\begin{equation}
    \psi_{n}(z) = \sqrt{\frac{2c \,n!}{(n+3)!}}\,(cz)^{7/2}\,e^{-c^2z^2/2} \,L_n^{3}(c^2z^2)\,, 
    \label{WFa}
\end{equation}
with associated mass spectrum given by 
\begin{equation}
    M_{n}^2 = 4 c^2 \left(n+2 \right)\,.
    \label{MSa}
\end{equation}

However, if we choose the full geometry computed from solving the EPD system, with a non-zero $A(z)$, the equation to be solved is given by 
\begin{gather}
%-\partial_z^2 \psi_n(z) + \left[\left(\partial_z \left[\frac{1}{2} \log\left(\frac{e^{-c^2z^2}}{z}\right)\right]\right)^2
%+ \partial_z^2 \left[\frac{1}{2} \log\left(\frac{e^{-c^2z^2}}{z}\right)\right] +\frac{8 \,e^{2A(z)}}{z^2} \right] \psi(z) = m_n^2 \psi(z) \nonumber \\ 
%\Leftrightarrow 
-\partial_z^2 \psi_n(z) + \left[\frac{3}{4z^2} + c^4z^2 + \frac{8 \,e^{\frac{-az^2}{4}}}{z^2} \right] \psi_n(z)  = m_n^2 \psi_n(z)  \,.
    \label{Fulleq}
\end{gather}

Unlike (\ref{harmonic}), it was not possible to find a closed-form solution to~(\ref{Fulleq}), consequently, one has to integrate it numerically. 

\begin{figure}[h!]
    \centering
    \includegraphics[width=0.75\textwidth]{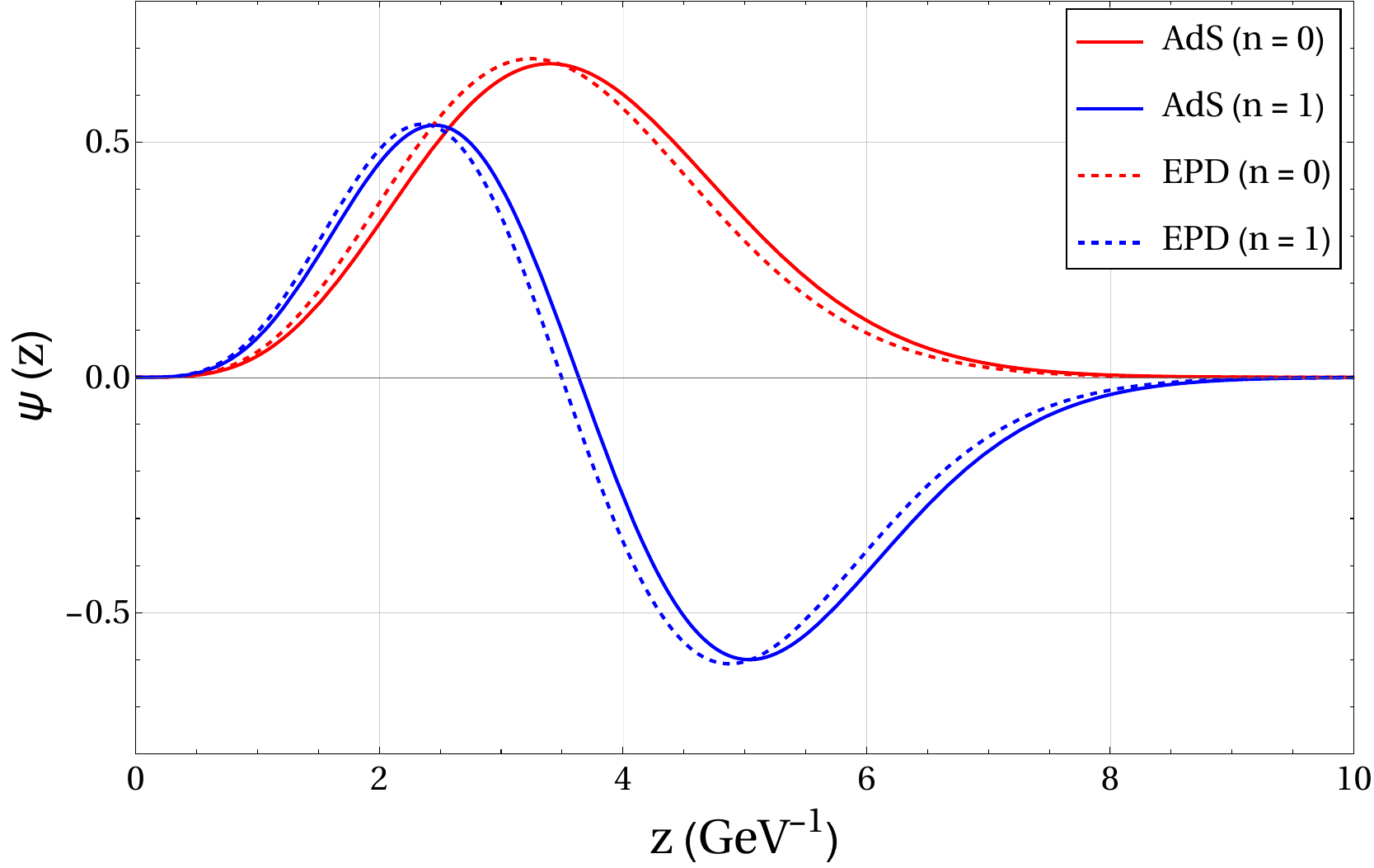}
    \caption{\small Wavefunctions at zero temperature in the AdS geometry, Eq.~(\ref{harmonic}), and the EPD geometry, Eq.~(\ref{Fulleq}).}
    \label{fig:WF0andA}
\end{figure}

\begin{figure}[h!]
    \centering
    \includegraphics[width=0.75\textwidth]{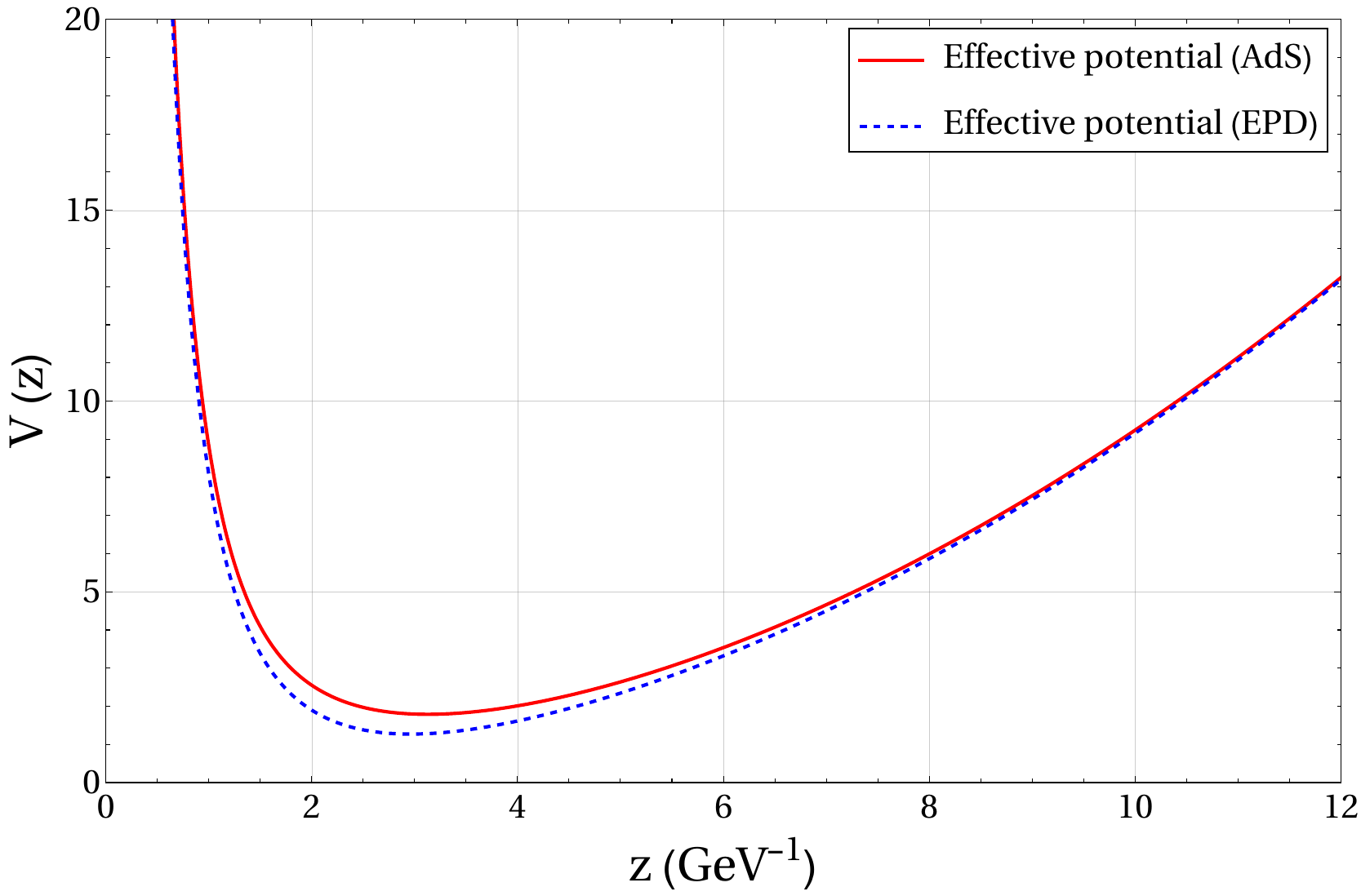}
    \caption{\small Effective potential~(\ref{Potential_T0}) at zero temperature in the AdS and EPD geometries.}
    \label{fig:V0andA}
\end{figure}

Equipped with the analytical solution of Eq.~(\ref{harmonic}), given in Eq.~(\ref{WFa}), we solve Eq.~(\ref{Fulleq}) numerically as a singular Sturm--Liouville eigenvalue problem by means of a shooting method. Near the asymptotic boundary, the effective potential behaves as $V(z)\sim 35/(4z^2)$, so that a Frobenius analysis, which can be seen in more detail in Section~\ref{sec43}, yields the two independent behaviors $\psi(z)\sim z^{7/2}$ and $\psi(z)\sim z^{-5/2}$. Normalizability selects the regular branch, $\psi(z)\sim z^{7/2}$, which is used to set the initial conditions at a small ultraviolet cutoff $z=\epsilon$. The equation is then integrated toward the infrared while varying the eigenvalue \(m_n^2\) until the normalizable condition $\psi_n(z_{\rm max})\simeq0$ is satisfied at a sufficiently large cutoff $z_{\rm max}$. \texttt{Mathematica} was used for the numerical evaluations. The resulting numerical wavefunctions for the EPD geometry, together with the analytical AdS solutions, are displayed in Fig.~\ref{fig:WF0andA}, while the corresponding effective potentials are compared in Fig.~\ref{fig:V0andA}. The mass spectrum is given analytically by Eq.~(\ref{MSa}) in the AdS geometry and by the numerical Sturm--Liouville eigenvalues of Eq.~(\ref{Fulleq}) in the EPD geometry, with the corresponding results collected in Table~\ref{thybrid}.

\begin{table}[h!]
\centering
\begin{tabular}{||c||c|c|c|c||}
    \hline\hline
    \multicolumn{5}{|c|}{Hybrid meson $\pi_1$, $I^G (J^{PC}) = 1^-(1^{-+})$} \\
    \hline
    $n$ & State & $m_{\text{exp}}$ (\text{MeV}) & $m_{\text{th}}$ (\text{MeV}) -- EPD & $m_{\text{th}}$ (\text{MeV}) -- AdS \\
    \hline
    0 & $\pi_1 (1400)$ & $1354 \pm 25$ & $1396.5$ & $1555.6$ \\\hline
    1 & $\pi_1 (1600)$ & $1660 \substack{+15 \\ -11}$ & $1796.3$ & $1905.3$ \\\hline
    2 & $\pi_1 (2015)$ & $2014 \pm 20 \pm 16$ & $2117.1$ &$2200.0$ \\\hline
    3 & ? & ? & $2392.8$ & $2459.7$ \\
    \hline\hline
    \multicolumn{5}{||c||}{Experimental linear fit: $m_n^2 = 1.11\,(n+1.59)$, $R^2 = 0.99913$} \\\hline
    \multicolumn{5}{||c||}{EPD geometry linear fit: $m_n^2 = 1.26\,(n + 1.56)$, $R^2 = 0.999996$} \\\hline
    \multicolumn{5}{||c||}{AdS geometry curve: $m_n^2 = c^2\,(4n+ 8)$} \\\hline
\end{tabular}
\caption{\footnotesize Regge trajectory of the $\pi_1$ hybrid meson and mass spectra for an AdS and a fully backreacted EPD geometry. Experimental data are read from PDG \cite{ParticleDataGroup:2024cfk}.
\label{thybrid}}
\end{table}

It is also interesting to compare the effective potential, seen in Eq.~(\ref{Potential_T0}), in both geometries. This is shown in Fig.~\ref{fig:V0andA}.

\begin{figure}[h!]
    \centering
    \includegraphics[width=0.75\textwidth]{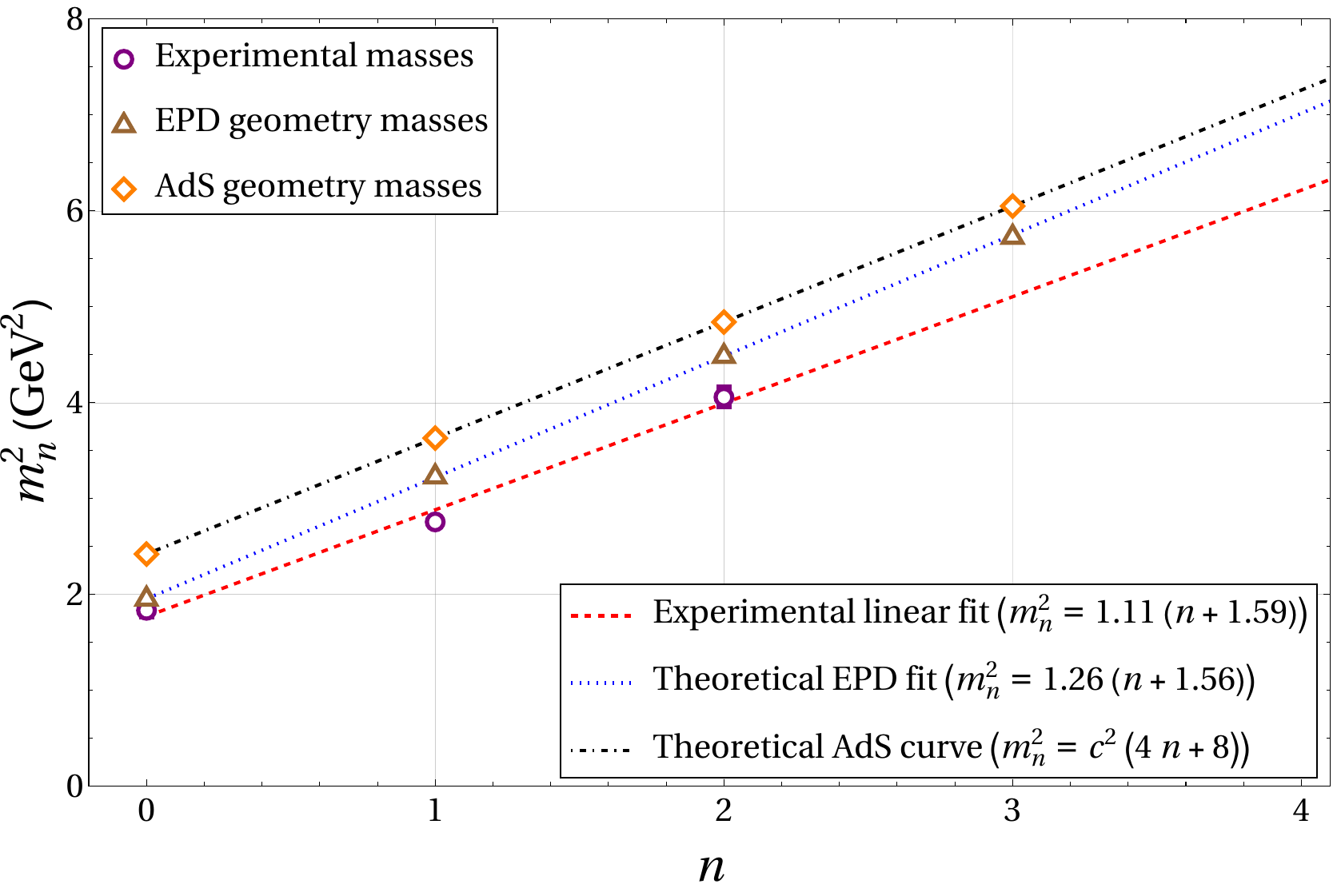}
    \caption{\small Regge trajectories for the hybrid meson.}
    \label{fig:Regge}
\end{figure}

For completeness, we also plot the experimental Regge trajectory for the hybrid meson and the obtained analytical and numerical masses in Fig.~\ref{fig:Regge}. 

From Figs.~\ref{fig:WF0andA} and \ref{fig:V0andA}, we immediately note that the wavefunction shapes and effective potentials are quite similar to each other for the studied cases. However, a noticeable change in the mass spectra is observed, as seen in Table~\ref{thybrid}. In fact, the new estimates are closer to the experimental values.
Importantly, we therefore conclude that a mathematically self-consistent EPD model changes the physical prediction for the mass spectra. This differs from what was observed in the case of an EMD model~\cite{Toniato:2025gts}, in which the mass spectra agree with a standard non-backreacting AdS (or AdS-Schwarzschild, in the finite-temperature case) geometry. The introduction of a massive (Proca) field leads to an $e^{-\frac{a}{4}z^2}$ correcting term to the equation of motion.

\section{The spectral functions in the EPD model}
\label{sec4}

In this section, the thermodynamic properties of the considered black hole system will be discussed, and the spectral functions of the hybrid meson will be computed.

\subsection{Equation of motion at finite temperature}

To study the system at finite temperature, we first determine the blackening function $g(z)$ for the selected warp-factor profile $A(z)=-\frac{a}{8}z^2$. It follows from Eq.~(\ref{metriceq}) that, when $\mu=0$ (so that $H_t=0$), the function $f(z)$ does not enter the equation for $g(z)$. The corresponding differential equation and boundary conditions are
\begin{equation}\label{ODE}
    g'' + g' \left(  3A' -\frac{3}{z} \right) = 0, \quad \{g(0) = 1, \, g(z_h) = 0\} \,,
\end{equation}
By integrating (\ref{ODE}), we obtain
\begin{equation}\label{g_T_mu_0}
g(z) = 1 + \frac{e^{\frac{3 a z^2}{8}} \left(8 - 3 a z^2\right) - 8}{e^{\frac{3 a z_h^2}{8}} \left(3 a z_h^2-8 \right) + 8}\,.
\end{equation}

The black hole temperature is calculated as
\begin{equation} \label{BH_T_mu0}
T(z_h) = -\frac{g'(z_h)}{4 \pi} = \frac{9  z_h^3 a^2 e^{\frac{3 a z_h^2}{8}}}{16 \pi \left[8 + e^{\frac{3 az_h^2}{8}} \left(3 a z_h^2-8\right)\right]}\,.
\end{equation}
To obtain the equation of motion for the perturbation in a black hole background, with a generic blackening function $g(z)$, we reinstate the $g(z)$ factors in the equation of motion~(\ref{EOM_2}) and obtain
\begin{equation}\label{FT1}
\frac{z e^{-A(z)}}{L f(z)}\partial_z\left[\frac{L e^{A(z)}f(z)g(z)}{z}\partial_z H_\mu\right]+\frac{1}{g(z)}\Box H_\mu-\frac{m_5^2L^2e^{2A(z)}}{z^2}H_\mu=0\,,
\end{equation}

In this and following computations, we proceed in the rest frame, where the four-momentum is given by $q^\mu = (\omega, 0, 0, 0)$. Employing the $ H_z = 0$ condition again, we consider plane wave solutions of the form
\begin{equation}
H^\mu (z, x^\nu) = e^{i q_\alpha x^\alpha} H^\mu(z, \omega) = e^{-i \omega t} H^\mu(z, \omega)\,.
\end{equation}
By reapplying (\ref{fourier}) and \eqref{B_def}, we transform the finite temperature equation of motion into
\begin{equation} \label{EOM_T}
-\partial_{r_*}^2 \psi(z) + V(z) \psi(z) = \omega^2 \psi(z)\,,
\end{equation}
where the tortoise coordinate $r_*$, implicitly defined by
\begin{equation}\label{tort}
\partial_{r_*} = -g(z) \partial_z\,,
\end{equation}
was introduced to transform (\ref{FT1}) into a Schr\"odinger-like equation. The associated effective potential is given by
\begin{equation} \label{Potential_T}
V(z) = \left(\partial_{r_*} \left[\frac{1}{2} \log\left(\frac{L e^{A(z)} f(z)}{z}\right)\right]\right)^2
+ \partial_{r_*}^2 \left[\frac{1}{2} \log\left(\frac{L e^{A(z)} f(z)}{z}\right)\right] + g(z)\frac{m_5^2L^2e^{2A(z)}}{z^2} \,.
\end{equation}
where $ \xi $ is the same function as defined in Eq.~(\ref{B_def}). With the equation of motion formulated at finite temperature, we can calculate the spectral functions in the EPD model. 

A further remark concerns the transversality condition employed in deriving the finite-temperature equation of motion~(\ref{EOM_T}). Since the bulk field is massive, this condition should not be interpreted as a gauge choice in the Maxwell sense. Rather, it specifies the transverse sector of the Proca dynamics. For nonvanishing spatial momentum, $\vec{q}\neq 0$, the momentum vector selects a preferred spatial direction and the fluctuations decompose into transverse and longitudinal channels, which generally obey different equations of motion and may therefore lead to distinct spectral functions, as discussed in Ref.~\cite{Dudal:2015kza}. In the present work, however, we restrict ourselves to the particle rest frame, $\vec{q}=0$, for which this distinction disappears and the transverse and longitudinal effective potentials coincide. The dynamics then reduce to the single equation of motion~(\ref{EOM_T}) used throughout the numerical analysis.

\subsection{Black hole thermodynamics and the parameter \texorpdfstring{$a$}{a}} \label{sec42}

Before we take on the task of computing the spectral functions, we first investigate the thermodynamic properties of the system and the determination of the parameter $a$. 

\begin{figure}[h!]
    \centering
    \includegraphics[width=0.75\textwidth]{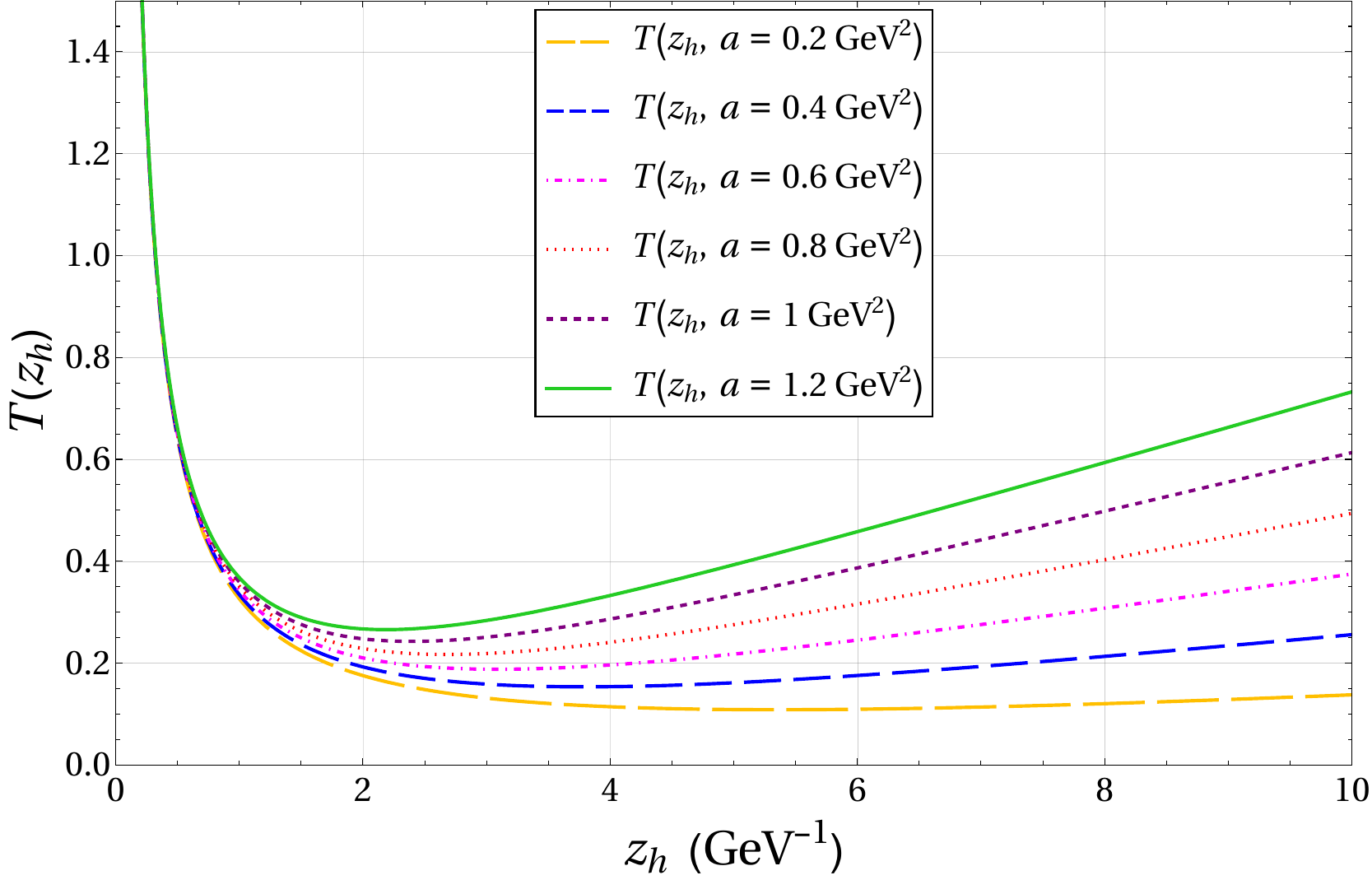}
    \caption{\small  The Hawking temperature $T$, given by~(\ref{BH_T_mu0}), as a function of horizon radius $z_h$ for various values of $a$.}
    \label{fig:BHtempA}
\end{figure}

Fig.~\ref{fig:BHtempA} displays the Hawking temperature $T$, computed in~(\ref{BH_T_mu0}) for the resulting metric~(\ref{g_T_mu_0}), as a function of the horizon position $z_h$ for various values of the parameter $a$. 

In this case, the thermal-AdS phase is available for all temperatures. Nevertheless, there exists a minimum temperature below which no black hole solution is present. For temperatures above this minimum, two branches of black holes appear: a large black hole and a small one. The large black hole branch (small $z_h$) exhibits a temperature that decreases as the horizon radius $z_h$ grows, and it is thermodynamically stable since its specific heat is positive. By contrast, the small black hole branch (large $z_h$) shows a temperature that increases with the horizon radius, and it is thermodynamically unstable due to its negative specific heat. We see that this effect is increasingly pronounced as $a$ gets larger. 

The black hole entropy is given by the usual Bekenstein–Hawking formula $S=\mathcal{A}_h/4G_5$, where $\mathcal{A}_h$ is the horizon area. The induced spatial metric, $t=\text{constant}$, for our metric Ans\"atze~(\ref{ansatz_metric}), on the horizon slice ($z=z_h$) is given by
\begin{equation}
    \gamma_{ij} = \frac{L^2 e^{2A(z_h)}}{z_h^2} \delta_{ij}\,,
\end{equation}
implying that 
\begin{equation}
\sqrt{\gamma}= \frac{L^3 e^{3A(z_h)}}{z_h^3}\,,
 \end{equation}
where $\gamma=\det{\gamma_{ij}}$. So we conclude that
\begin{equation}
    \mathcal{A}_h = \int d^3x \sqrt{\gamma} = V_3 \frac{L^3 e^{3A(z_h)}}{z_h^3}\,,
\end{equation}
yielding 
\begin{equation}
S = \frac{V_3}{4G_5} \frac{L^3 e^{3A(z_h)}}{z_h^3}\,.
    \label{BHentropy}
    \end{equation}

Since both the entropy and temperature are functions of the horizon position $z_h$, the thermodynamic relation $dF=-S\,dT$ can be written as
\begin{equation}
\frac{dF}{dz_h}=-S(z_h)\frac{dT(z_h)}{dz_h}\,,
\end{equation}

Choosing the integration constant such that $F(z_h\to\infty)=0$, we obtain
\begin{equation}
F(z_h)=\int_{z_h}^{\infty} dx\, S(x) T'(x)=-S(z_h)T(z_h)-\int_{z_h}^{\infty} dx\, S'(x)T(x)\,.
\label{freeE}
\end{equation}

\begin{figure}[h!]
    \centering
    \includegraphics[width=0.75\textwidth]{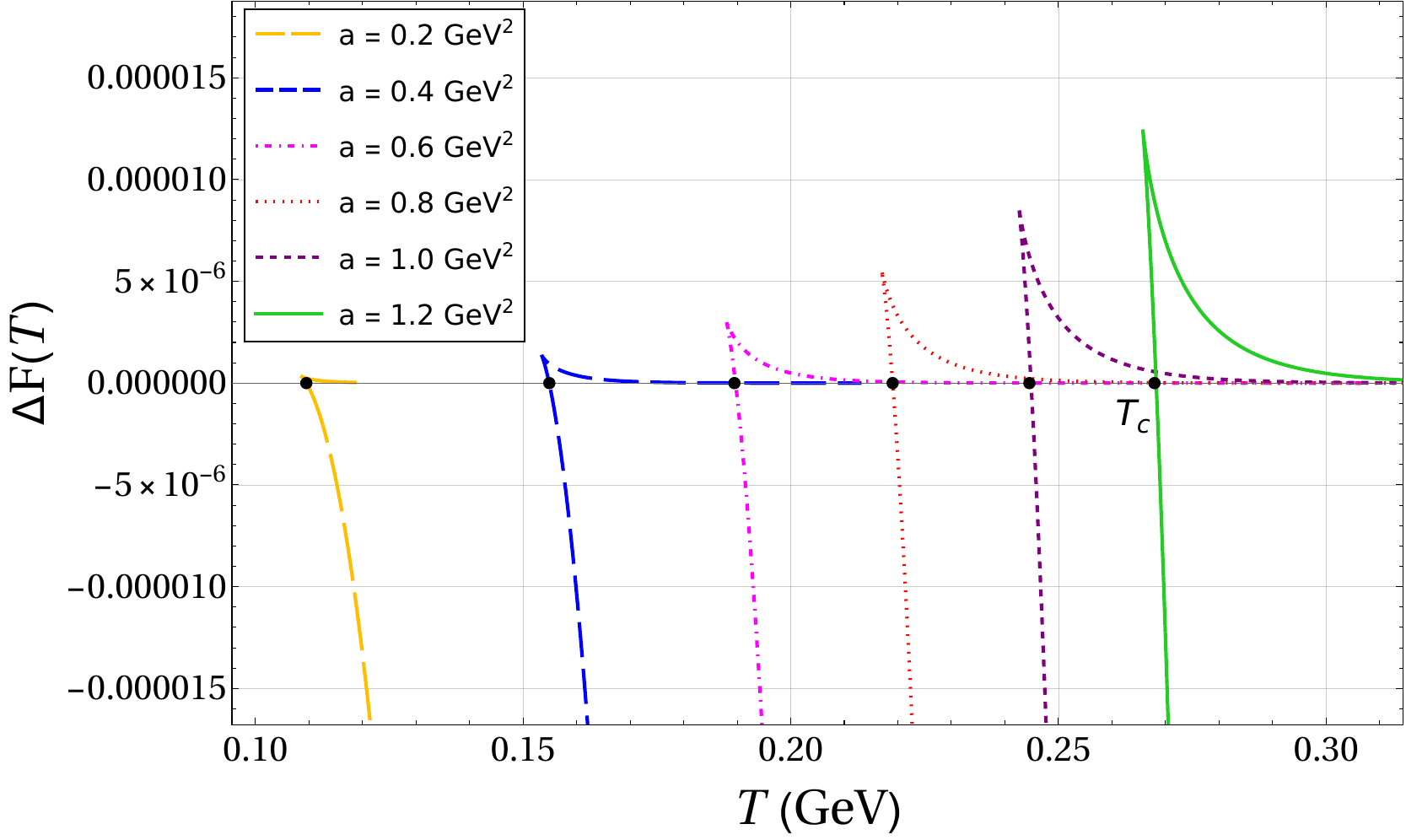}
    \caption{\small Free energy difference for various values of $a$.}
    \label{fig:BHfreeA}
\end{figure}

Here, the two phases considered are the thermal AdS and the black hole. Therefore, we can define the free energy difference as $\Delta F=F_{\text{BH}}-F_{\text{AdS}}$, for the computed free energy~(\ref{freeE}). The $\Delta F$ diagram is displayed in Fig.~\ref{fig:BHfreeA}.

For the curves shown in Fig.~\ref{fig:BHfreeA}, we adopt a normalization in which the thermal-AdS free energy is set to zero. The free-energy difference $\Delta F$ is then evaluated for a range of values of $a$. We observe that, upon varying the temperature, the preferred phase switches between the large black hole branch and the thermal-AdS background. In particular, the large black hole yields the lowest free energy at high temperatures, while thermal-AdS is favored at low temperatures. This indicates a first-order Hawking--Page transition separating the two phases. For instance, if $a=1~\text{GeV}^2$ we find the transition at $T_c=245~\text{MeV}$. From the boundary perspective, thermal AdS corresponds to the confined phase, whereas the black hole geometry represents the deconfined phase. Note, moreover, that the small black hole branch always lies above both the large black hole and thermal-AdS branches in free energy, so it constitutes the global maximum of the solutions and is thermodynamically disfavored at all temperatures. Accordingly, in the remainder of this work, we focus on the large black hole branch when discussing the finite-temperature behavior of the observables under consideration.

\begin{figure}[h!]
    \centering
    \includegraphics[width=0.75\textwidth]{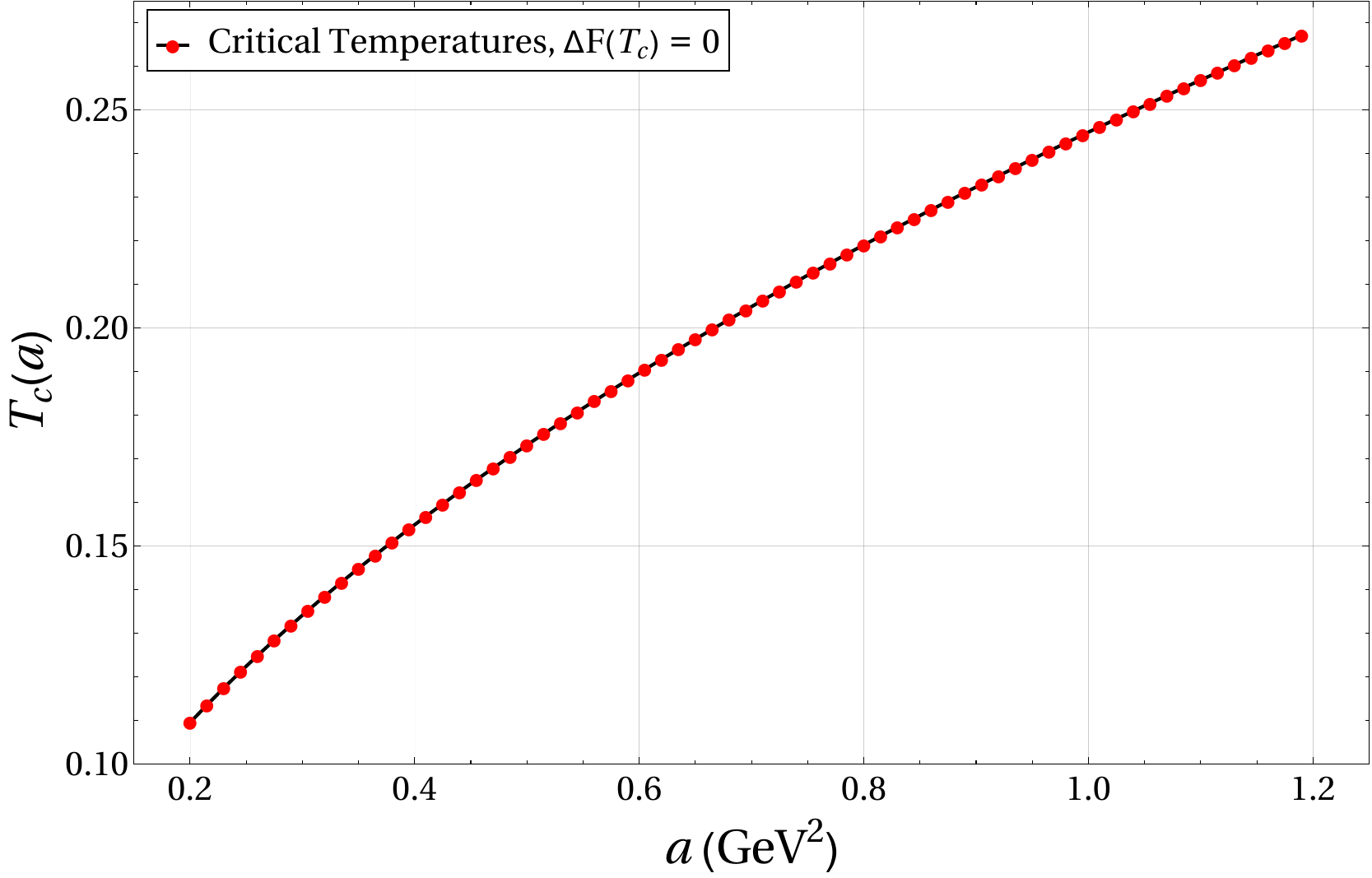}
    \caption{Critical Hawking--Page temperature $T_c$, defined by $\Delta F(T_c)=0$, as a function of $a$}
    \label{fig:CriticalA}
\end{figure}

If we apply the free energy difference computation to the range $a\in[0.2~\text{GeV}^2,1.2~\text{GeV}^2]$ and extract their respective critical temperature Hawking--Page transition, we find the curve presented in Fig.~\ref{fig:CriticalA}.

We readily notice that the critical temperature rises nonlinearly with $a$. The QCD phase transition temperature is set around $150~\text{MeV}$ in the lattice estimate of \cite{Steinbrecher:2018phh}. For this critical temperature, $a\sim0.395~\text{GeV}^2$. In the case where $A(z)\ne0$, we, therefore, use $a=0.395\text{GeV}^2$ in this work. 

\subsection{Spectral functions in a non-backreacting AdS-Schwarzschild geometry}\label{sec43}

In a similar spirit to what was done in Section~\ref{sec3}, we set out to first calculate the finite temperature effects on hybrid mesons according to a non-backreacting AdS-Schwarzschild geometry. To do so, we take the appropriate limits of~(\ref{g_T_mu_0}) and~(\ref{BH_T_mu0}) as $a\to 0$:
\begin{equation}
g_{\text{AdS-Schwarzschild}}(z) = \lim_{a \to 0} \left( 1 + \frac{e^{\frac{3 a z^2}{8}} \left(8 - 3 a z^2\right) - 8}{e^{\frac{3 a z_h^2}{8}} \left(3 a z_h^2-8 \right) + 8} \right) =1 - \frac{z^4}{z_h^4}\,, 
\label{gAdS}
\end{equation}
\begin{equation}
T_{\text{AdS-Schwarzschild}}(z_h) = \lim_{a \to 0} \left(  \frac{9  z_h^3 a^2 e^{\frac{3 a z_h^2}{8}}}{16 \pi \left[8 + e^{\frac{3 az_h^2}{8}} \left(3 a z_h^2-8\right)\right]} \right) = \frac{1}{\pi z_h}\,.
\label{TAdS}
\end{equation}

In this work, we follow the prescription of Refs. \cite{Jena:2024cqs,Toniato:2025gts} to evaluate the spectral functions. In a standard non-dynamical soft-wall setup, where Einstein's equations need not be satisfied, the factor $1/g(z)$ is often integrable in closed form, yielding an explicit expression for the tortoise coordinate $r_*$ as it is derived from \eqref{tort}. 

As an example, for the blackening function given above in Eq.~(\ref{gAdS}), one finds the following after integration
\begin{eqnarray}
	r_*^{\text{AdS}}(z) = \frac{z_h}{2} \left[\frac{1}{2}\log\left({\frac{z_h-z}{z_h+z}}\right) -\arctan\left({\frac{z}{z_h}}\right) \right]\,.
    \label{TortA}
\end{eqnarray}

We readily note that $ g_{\text{AdS-Schwarzschild}}(z_h) = 0 $, which implies that the equations of motion simplify to a free-particle form near the horizon. Consequently, the solutions close to the horizon can be expressed as ingoing and outgoing plane waves traveling into and out of the black hole interior as
\begin{eqnarray}
    \psi \sim \mathbb{C} \,e^{-i\omega r_*^{AdS}} + \mathbb{D} \,e^{i\omega r_*^{AdS}}\,.
\end{eqnarray}
where $ \mathbb{C} $ and $ \mathbb{D} $ are constants. 

In the case studied here, set by the Schr\"odinger-like equation~(\ref{EOM_T}) with effective potential~(\ref{Potential_T}) for the metric~$(\ref{gAdS})$, however, a simpler and equivalent expansion can be found. 

In this vein, the solutions to the equation of motion~(\ref{EOM_T}) near the horizon can be expanded in a power series as
\begin{eqnarray}\label{powers}
    \psi_{\pm} = \left(1-\frac{z}{z_h}\right)^{\pm i\omega z_h/4} \left[ a_0^{\pm} + a_1^{\pm} \left(1-\frac{z}{z_h}\right)+ a_2^{\pm} \left(1-\frac{z}{z_h}\right)^2 + \dots \right]\,,
\end{eqnarray}
where we set $a_0^{\pm} = 1$ for normalization purposes. By substituting this expansion into the equation of motion~(\ref{EOM_T}), we can determine the coefficients $a_1^{\pm}$ and $a_2^{\pm}$. These coefficients depend on the frequency $\omega$ and momentum $q$. For subsequent computations, the solutions $\psi_{\pm}$ are truncated at the second-order term, $\left(1-\frac{z}{z_h}\right)^2$. The computed coefficients $a_1^-$ and $a_2^-$ for the expansion (\ref{powers}) are given by
\begin{align}
    a_1^- &= -\frac{i}{8}  \left[3 \omega  z_h+\frac{-4 z_h^2 \left(4 c^2+q^2\right)-40}{\omega  z_h+2 i}\right]\,,\\
    a_2^- &= \frac{1}{128 \,(\omega  z_h+2 i) (\omega  z_h+4 i)} \Biggl[-4480+z_h \biggl(128 i c^4 z_h^3 (\omega  z_h+4 i)-32 c^2 z_h \nonumber \\
    &\times \left(4 q^2 z_h^2+\omega 
   z_h (-3 \omega  z_h+4 i)+32\right)-16 q^4 z_h^3 +8 q^2 z_h  [-64+3 \omega  z_h (\omega 
   z_h+4 i)] \nonumber\\ 
   &+\omega  \{\omega  z_h [432-\omega  z_h (9 \omega  z_h+74 i)]+1600 i\}\biggl)\Biggl],
\end{align}
whereas the coefficients $a_{1,2}^+$ are quite similar to their counterparts shown above and are not actually employed in Eq.~(\ref{green}) to compute the numerical spectral functions.

The ingoing and outgoing solutions form a basis for any wave function that may be considered. Consequently, normalizable and non-normalizable solutions can be expressed as linear combinations of the ingoing and outgoing solutions
\begin{align}
    \psi_{2} &= \mathbb{C}_2 \psi_- + \mathbb{D}_2 \psi_+ \,, \\
    \psi_{1} &= \mathbb{C}_1 \psi_- + \mathbb{D}_1 \psi_+ \,.
\end{align}

Additionally, in the limit $ z \to 0 $, the leading term in the power series solution is assumed to take the form $ z^\beta $, following the prescription for a Frobenius expansion near $ z = 0$. The asymptotic analysis of Eq.~(\ref{EOM_T}) yields the following indicial equation
\begin{equation}
    -\beta^2 + \beta + \frac{35}{4} = 0\,,
\end{equation}
which has solutions $ \beta = -\frac{5}{2} $ and $ \beta = \frac{7}{2}$. When solving the perturbation equation by integrating from the boundary toward the horizon, one must also fix the boundary value of $\partial_z\psi_i$ to ensure a well-posed numerical procedure. Determining this first derivative at the holographic boundary entails analyzing the asymptotic behavior of $\psi_i$ through a Frobenius expansion. Because the indicial equation yields two roots that differ by an integer, the larger exponent, $\beta=\frac{7}{2}$, gives a regular power-series solution, while the smaller exponent, $\beta=-\frac{5}{2}$, has an addition of a logarithmic term. These two independent solutions may be written as
\begin{equation}
    \begin{aligned}
        \psi_1(z) &= z^{7/2} \sum_{k=0}^{\infty} a_k z^{2k} \,, \\
        \psi_2(z) &= K \ln(z) \,\psi_1(z) + z^{-5/2} \sum_{k=0}^{\infty} b_k z^{2k} \,.
    \label{power}
    \end{aligned}
\end{equation}

The explicit forms of $ a_k $ and $ b_k $ can be determined by substituting the series solutions into Eq.~(\ref{EOM_T}) and matching terms order by order. For normalization, we choose $a_0 = b_0 = 1$. The coefficients of these expansions depend on $ \omega $ and momentum $q$. In this work, the presented expansion (\ref{power}) for $ \psi_1(z) $ and $ \psi_2(z) $ was truncated at $k=2$. Accordingly, for the considered equation of motion (\ref{EOM_T}), the computed coefficients are given by
\begin{align}
    a_1 &= \frac{1}{16} \left(q-\omega \right)\left(q+\omega \right) \,, \\
    a_2 &= \frac{1}{640} \left[16 c^4 + \frac{384}{z_h^4} + \left(q^2-\omega^2\right)^2 \right] \,,  \\
    b_1 &= 0 \,,  \\
    b_2 &= -\frac{c^4}{8} \,,  \\
    K &= \frac{1}{48 \,z_h^4} \left[32 c^2- 8q^2 + c^4 z_h^4 (\omega^2-q^2)\right] \,.
 \label{power1}
\end{align}
in which the coefficient $b_1$ can be chosen freely, as shown in \cite{Mamani:2022qnf}, and here is taken to be null.

These solutions, starting near the boundary, also form a basis for constructing wave functions, which can be written as
\begin{align}
    \psi_{+} &= \mathcal{A}_+ \psi_2 + \mathcal{B}_+ \psi_1 \,, \\
    \psi_{-} &= \mathcal{A}_- \psi_2 + \mathcal{B}_- \psi_1 \,.
\end{align}

Both the near-boundary and near-horizon expansions provide a basis in which wave functions can be expressed as linear combinations. From the relations between $ \psi_{\pm} $, $ \psi_1 $, and $ \psi_2 $, it follows that the coefficients satisfy
\begin{gather}
 \begin{pmatrix} \mathcal{A}_{-} & \mathcal{B}_{-} \\ \mathcal{A}_{+} & \mathcal{B}_{+} \end{pmatrix}
 =
  \begin{pmatrix} \mathbb{C}_{2} & \mathbb{D}_{2} \\ \mathbb{C}_{1} & \mathbb{D}_{1} \end{pmatrix}^{-1}\,.
\end{gather}

Our primary interest lies in the computation of the spectral functions. Within this approach, this requires determining the retarded Green's function and subsequently extracting its imaginary part  \cite{Mamani:2022qnf,Jena:2024cqs,Toniato:2025gts}, thereby effectively implementing the real-time AdS/CFT dictionary as set out in \cite{Son:2002sd,Policastro2002Oct}.

Generally, the coefficients depend on the momentum $q$, as seen in the computed expansions above. However, as previously stated in this work, we take $q=0$.

The retarded Green's function is obtained by numerically integrating the equations of motion on a regular lattice from a point close to the boundary, $z=\epsilon$, up to a point near the horizon, $z=z_h-\epsilon$. The required initial data are taken from the $z=0$ near-boundary solution, namely $\psi_1(\epsilon)$ as determined via Frobenius analysis, together with its derivative $\partial_z\psi_1(\epsilon)$. Likewise, the second independent near-boundary mode, $\psi_2(\epsilon)$, along with $\partial_z\psi_2(\epsilon)$, provides the initial conditions used to generate the other independent solution of the equations of motion within the same numerical procedure.

As a result, two numerical solutions for $ \psi $ are obtained. These functions are then evaluated near the horizon, $ z_h - \epsilon $, where the following relationship is constructed
\begin{gather}
\begin{pmatrix} \psi_j(z_h - \epsilon) \\ \partial_z \psi_j(z_h - \epsilon) \end{pmatrix}
=
\begin{pmatrix} \psi_-(z_h - \epsilon) & \psi_+(z_h - \epsilon) \\ \partial_z \psi_-(z_h - \epsilon) & \partial_z \psi_+(z_h - \epsilon) \end{pmatrix}
\begin{pmatrix} \mathbb{C}_j \\ \mathbb{D}_j \end{pmatrix}\,,
\label{coef}
\end{gather}
where $j=1,2$ correspond to the numerical solutions obtained by integrating the equations of motion from the boundary, while $\psi_{\pm}$ denote the asymptotic ingoing and outgoing solutions near the horizon, given by the expansion in Eq.~(\ref{powers}) and its corresponding coefficients. All quantities are evaluated at $z_h-\epsilon$. The coefficients $C_j$ and $D_j$ are obtained by inverting the above matrix.

Combining this with the coefficient relations in Eq.~(\ref{coef}), we arrive at
\begin{eqnarray} \label{green}
    \frac{\mathcal{B}_-}{\mathcal{A}_-} = - \frac{\mathbb{D}_2}{\mathbb{D}_1} = - \frac{\partial_z \psi_-(z_h - \epsilon) \psi_2(z_h - \epsilon) - \psi_-(z_h - \epsilon) \partial_z \psi_2(z_h - \epsilon)}{\partial_z \psi_-(z_h - \epsilon) \psi_1(z_h - \epsilon) - \psi_-(z_h - \epsilon) \partial_z \psi_1(z_h - \epsilon)}\,,
\end{eqnarray}

This provides a stable numerical method for calculating the ratio $ \frac{\mathcal{B}_-}{\mathcal{A}_-} $, which is sufficient to compute the retarded Green's function and, consequently, the spectral function. As demonstrated in Refs. \cite{Miranda:2009uw, Mamani:2022qnf}, the spectral function is directly related to $ \frac{\mathcal{B}_-}{\mathcal{A}_-} $ via the expression
\begin{eqnarray}
    \rho(\omega, q = 0) = - 2 \Im G^R(\omega, q = 0) \propto \Im \frac{\mathcal{B}_-}{\mathcal{A}_-}\,.
    \label{FormSpec}
\end{eqnarray}
where $ \rho(\omega, q = 0) $ represents the spectral function calculated numerically using the described prescription (with $ q = 0 $ set initially), and $ \Im $ denotes the imaginary part.

To evaluate $\psi_i(z_h-\epsilon)$ and $\partial_z\psi_i(z_h-\epsilon)$ appearing in the Green's-function expression~(\ref{green}), we numerically integrate the equation of motion over the interval $z\in[\epsilon,\,z_h-\epsilon]$. This is implemented in \texttt{Mathematica} using \texttt{NDSolve} with the \texttt{StiffnessSwitching} scheme, which we employ to handle the stiffness of the system, since explicit integrators such as the standard fourth-order Runge--Kutta or Euler methods were unable to yield stable and trustworthy results. In addition, the sizeable separation between the Frobenius exponents $\beta$ compels us to adopt a high accuracy goal, increasing the number of digits retained in the numerical evaluation.

The remaining ingredients in Eq.~(\ref{green}), namely $\psi_{\pm}(z_h-\epsilon)$ and $\partial_z\psi_{\pm}(z_h-\epsilon)$, are determined analytically from the Frobenius analysis and the corresponding power-series expansions.

The small positive parameter $ \epsilon $ functions as both the $\omega$-lattice spacing and the boundary cutoff. For the computations performed, a value of $ \epsilon = 0.01 \,\text{GeV}^{-1} $ was used. The value of $ z_h$, corresponding to a specific temperature $ T$, is determined by solving the transcendental equation (\ref{BH_T_mu0}) for $ z_h$ using Newton's method when taking the full geometry, $A(z)\ne0$. However, for $A(z)=0$ (AdS-Schwarzschild geometry), we have the simple relation $T=1/(\pi z_h)$.

\begin{figure}[h!]
    \centering
    \includegraphics[width=0.75\textwidth]{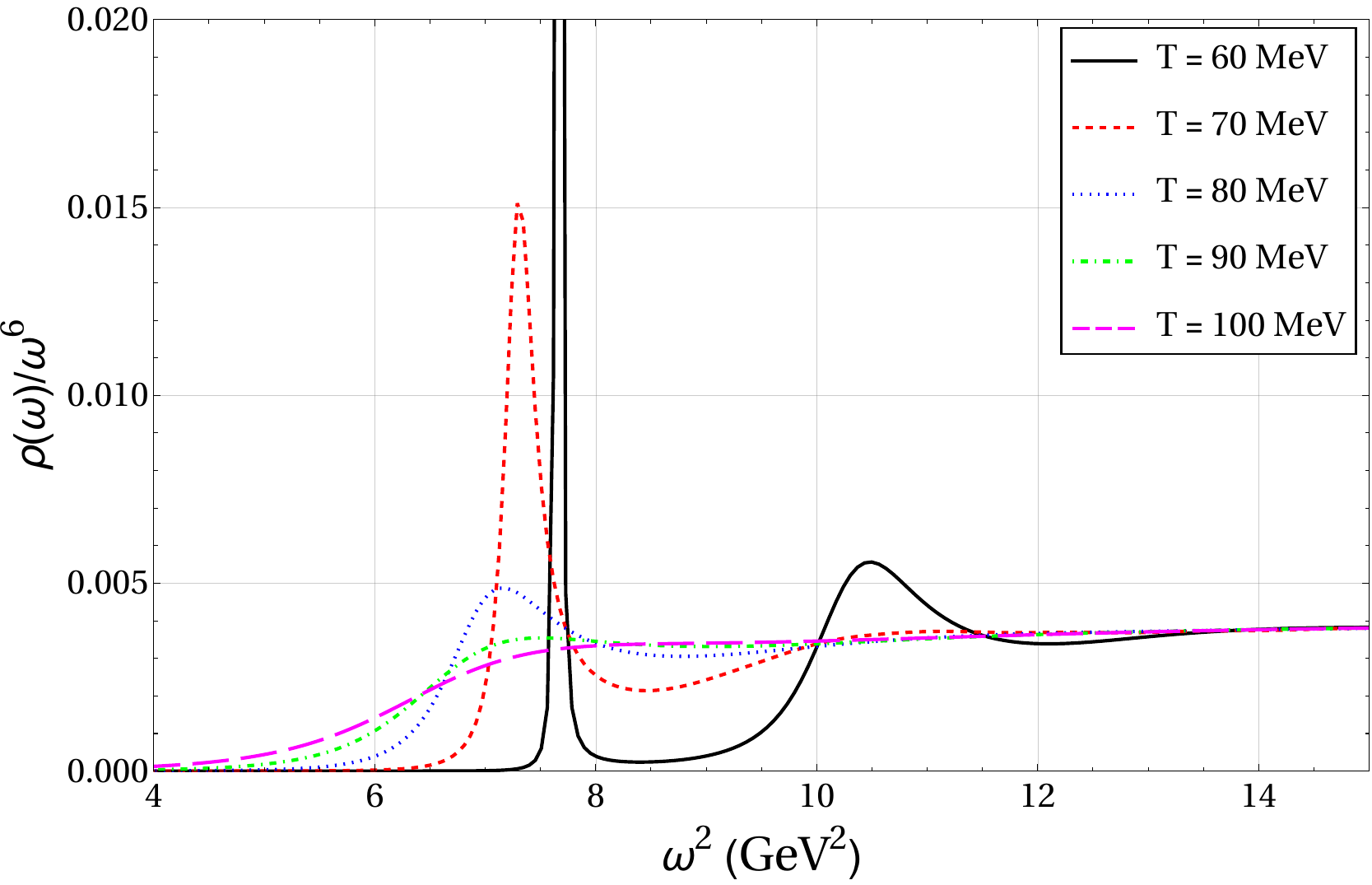}
    \caption{\small Spectral function for various values of temperature in the AdS-Schwarzschild geometry with $c=1~\text{GeV}$.}
    \label{fig:EDPSpec0}
\end{figure}

\begin{figure}[h!]
    \centering
    \includegraphics[width=0.75\textwidth]{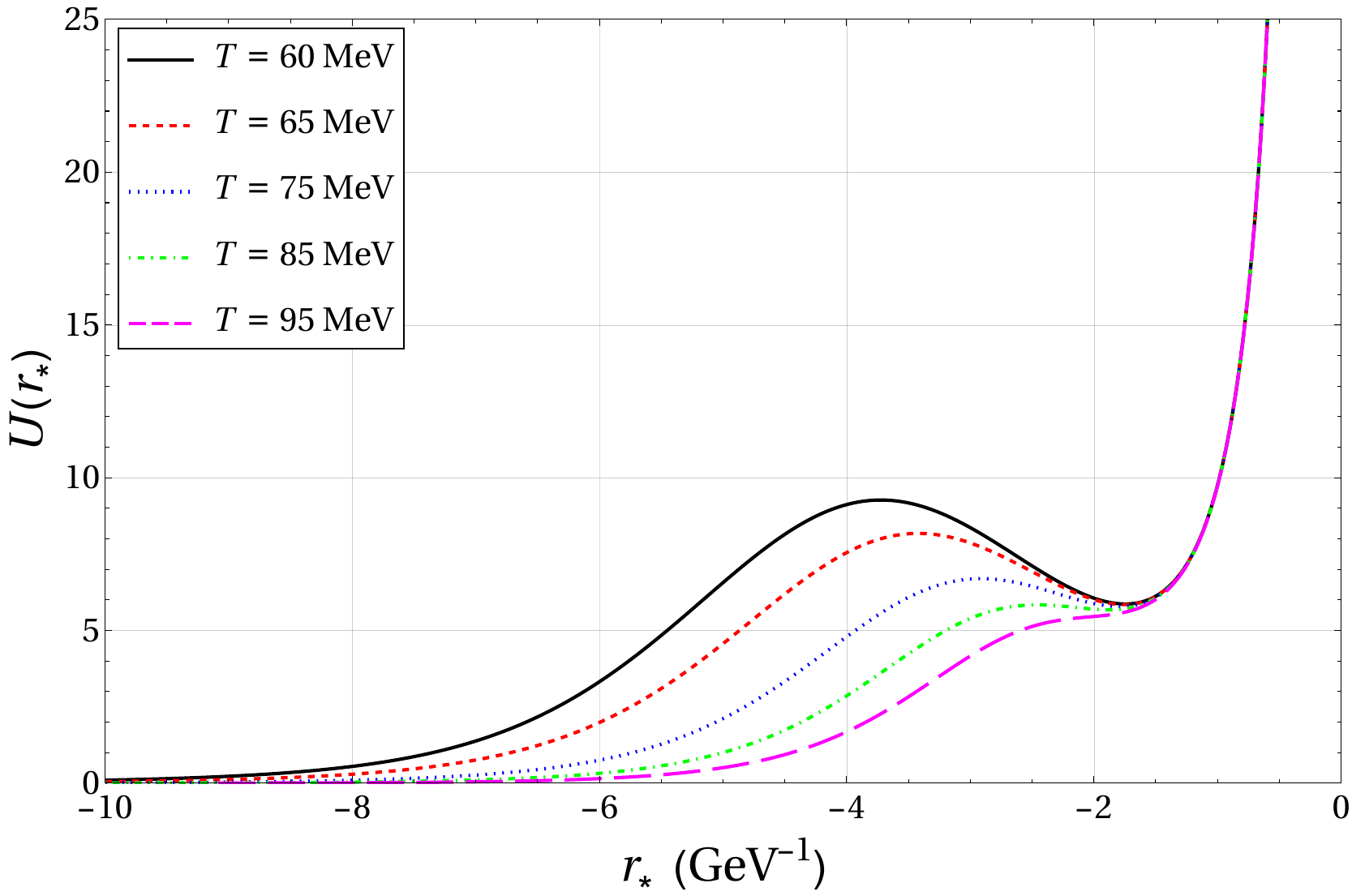}
    \caption{\small Effective potential for various values of temperature in the AdS-Schwarzschild geometry with $c=1~\text{GeV}$.}
    \label{fig:EPDU0}
\end{figure}

Equipped with the numerical procedure described above, we can compute the spectral functions for the non-backreacting AdS-Schwarzschild geometry. The corresponding spectral functions are shown in Fig.~\ref{fig:EDPSpec0} and the associated effective potentials in Fig.~\ref{fig:EPDU0} for a range of temperatures. The effective potentials are evaluated in the tortoise coordinate~(\ref{TortA}) variable.

As illustrated in Fig.~\ref{fig:EDPSpec0}, the spectral functions provide a clear signature of the thermal melting process. With increasing temperature, the pronounced resonance peaks associated with well-defined mesonic excitations progressively lose height and widen, until they are ultimately washed out into a smooth continuum. In physical terms, this broadening reflects the growth of thermal damping in the medium and signals that the would-be bound states cease to behave as long-lived quasiparticles, effectively dissolving into the hot environment.

A complementary way to see the same phenomenon is to inspect the temperature dependence of the corresponding effective potentials. As shown in Fig.~\ref{fig:EPDU0}, raising $T$ makes the potential well increasingly shallow, while the characteristic global minimum that supports localized (bound-state-like) solutions is gradually lifted and can disappear altogether. Once the well is sufficiently reduced, it can no longer sustain sharp resonant structures, in direct correspondence with the fading peaks observed in the spectral functions.

We computed the spectral functions and effective potentials with $c=1~\text{GeV}$. This is done to facilitate the comparison with~\cite{Bellantuono:2014lra}. We readily note that the results are equal, up to some normalization factor. Having reproduced the known AdS-Schwarzschild geometry result for hybrid mesons at finite temperature, we compute such quantities using the fully backreacted EPD geometry\footnote{During the preparation of this manuscript, we became aware of \cite{Zhao:2026bsc} where similar findings were reported, albeit not with a self-consistent (non-backreacted) model.}.

\subsection{Spectral functions in the fully backreacted geometry}

The actual numerical procedure followed here is the same as the one described above for the last case. However, there are significant nontrivial changes to the expansions as we take the full form of Eq.~(\ref{EOM_T}) in the fully backreacted EPD geometry. 

We note that $g(z_h)=0$, with $g$ defined by~($\ref{g_T_mu_0}$), so we can expand the solutions near the horizon as plane waves. As in the last case, it is possible to find a simpler and equivalent expansion. Accordingly, the near-horizon power expansion takes the form
\begin{eqnarray}\label{powers2}
    \psi_{\pm} = \left(1-\frac{z}{z_h}\right)^{\pm i\omega \chi(z_h)} \left[ a_0^{\pm} + a_1^{\pm} \left(1-\frac{z}{z_h}\right)+ a_2^{\pm} \left(1-\frac{z}{z_h}\right)^2 + \dots \right]\,,
\end{eqnarray}
where
\begin{eqnarray}
    \chi(z_h) = \frac{4 \left(3  a z_h^2+8 e^{-3 a z_h^2/8}-8\right)}{9 a^2 z_h^3}  \,,
\end{eqnarray}

As a sanity check, if we take the appropriate limit $a\to0$, we find
\begin{equation}
    \lim_{a\to0}\chi(z_h)= \frac{z_h}{4}\,.
\end{equation}
which gives the same expansion (\ref{powers}) found for the AdS-Schwarzschild case.

Again, we set $a_0^{\pm} = 1$. By substituting the expansion~(\ref{powers2}) into the equation of motion~(\ref{EOM_T}), we can determine the coefficients $a_1^{\pm}$ and $a_2^{\pm}$. We show below $a_1^{-}$ and $a_2^{-}$. For numerical computations, the expansion is truncated at second order.
\begin{align}
    a_1^- &= \frac{1}{6 a^2 z_h^3 \left[64 \omega +e^{3 a z_h^2/8} \left(-64 \omega +3 a z_h^2 \{8 \omega +3 i a z_h\}\right) \right]} \nonumber \\ 
          &\times i \Biggl[e^{3 a z_h^2/8} \biggl(27 a^4 z_h^6 \left(2 c^2 z_h^2+1\right)+24 a^2 q^2 z_h^4 \left(3 a z_h^2-8\right) - 9 i a^2 \omega
    z_h^3 \left(a z_h^2+4\right) \left(3 a z_h^2-8\right)\nonumber \\ 
          &- 8 \omega ^2 \left(8-3 a z_h^2\right)^2 \left[a z_h^2+4\right]\biggl)+8
   \biggl(24 a^2 q^2 z_h^4-9 i a^2 \omega  z_h^3 \left(a z_h^2+4\right) -16 \omega ^2 \left(a z_h^2+4\right) \nonumber \\
   &\times \left(3 a
   z_h^2-8\right) +1536 a^2 z_h^2 e^{-a z_h^2/4}+192 a^2 z_h^2 e^{a z_h^2/8} \left(3 a z_h^2-8\right) - 512 \omega ^2 e^{-3 a z_h^2/8} \nonumber \\
   &\times \left[a z_h^2+4\right] \biggl) \biggl] \,, \\
    a_2^- &= - \frac{288\, a^{-4} z_h^{-6} \left[32 \omega +e^{3 a z_h^2/8} \left(-32 \omega +3 a z_h^2 \{4 \omega +3 i a z_h\}\right)\right]^{-1} }{64 \omega
   +e^{3 a z_h^2/8} \left(-64 \omega +3 a z_h^2 (8 \omega +3 i a z_h)\right)} \nonumber \\
   &\times e^{-3az_h^2/4}\left(8388608 \,\omega ^4 \left(e^{3 a z_h^2/8}-1\right)^4-729 i \omega   a^9 z_h^{15} e^{3 a
   z_h^2/2}-\dots \right)\,.
\end{align}
where $a_2^-$ has a large number of terms, so we showed only a few of them. These expansions also form a basis for solutions.

The Frobenius analysis in the EPD geometry gives the same $\beta=7/2$ and $\beta=-5/2$ coefficients as the AdS-Schwarzschild geometry so we have the same form of expansion~(\ref{power}) near zero. However, the computed coefficients are somewhat more complicated, as seen below
\begin{align}
    a_1 &= \frac{1}{16} \left(q^2-\omega^2-2a \right) \,, \\
    a_2 &= \frac{1}{640} \left[16 c^4 + a^2\left(8+\frac{216}{8+e^{3az_h^2/8}\left(3az_h^2-8\right)}\right) + \left(q^2-\omega^2\right)^2 + 4a\left(\omega^2-q^2\right) \right] \,,  \\
    b_1 &= 0 \,,  \\
    b_2 &= -\frac{1}{32}\left(4c^2+a^2 \right) \,,  \\
    K &= \frac{1}{576} \Biggl[24a c^4 \!+\! 4a^3 \!-\! 12q^2c^4 \!-\! 3a^2q^2 \!+\! \frac{54 a^2 \left(4c^2+a-q^2 \right)}{8\!+\!e^{3az_h^2/8}\left(3az_h^2\!-\!8\right)} \!+\! 12 c^4\omega^2 \!+\!3 a^2\omega^2  \Biggl] \,.
 \label{power2}
\end{align}
$a_0=b_0=1$ was chosen. These expansions also form a basis for solutions. 

The same steps are applied to arrive at the spectral function formula~(\ref{FormSpec}) as the retarded Green's function is computed in the same way. 

One crucial difference, however, is that the EPD geometry, as seen in the thermodynamics Section~\ref{sec42}, only admits valid black hole solutions above a minimum value of temperature. There is no solution for the transcendental equation for $z_h$ defined by the temperature~(\ref{BH_T_mu0}) if $T$ is below such a temperature. Therefore, we are limited to studying the melting behavior of our system at $T\sim150~\text{MeV}$ and above, as this is the Hawking--Page transition temperature adopted in this work.

For the fully backreacted EPD geometry, the tortoise coordinate must be constructed from the corresponding blackening function in Eq.~(\ref{g_T_mu_0}), rather than from the AdS--Schwarzschild expression in Eq.~(\ref{TortA}). Following the prescription adopted in Ref.~\cite{Toniato:2025gts}, where the same blackening function was considered, we approximate
\begin{equation}
\frac{d r_*^{\rm EPD}}{dz}=-\frac{1}{g(z)}
\end{equation}
by performing a series expansion of $1/g(z)$ up to the same truncation order employed in \cite{Toniato:2025gts} and integrating the resulting expression, with the integration constant chosen such that $r_*^{\rm EPD}(0)=0$. The effective potential is then evaluated parametrically as $V\!\left[r_*^{\rm EPD}(z)\right]$. Consequently, the AdS--Schwarzschild expression~(\ref{TortA}) is used only in Sec.~\ref{sec43}, whereas the effective potentials in the fully backreacted geometry are expressed in terms of the EPD tortoise coordinate defined above.

\begin{figure}[h!]
    \centering
    \includegraphics[width=0.75\textwidth]{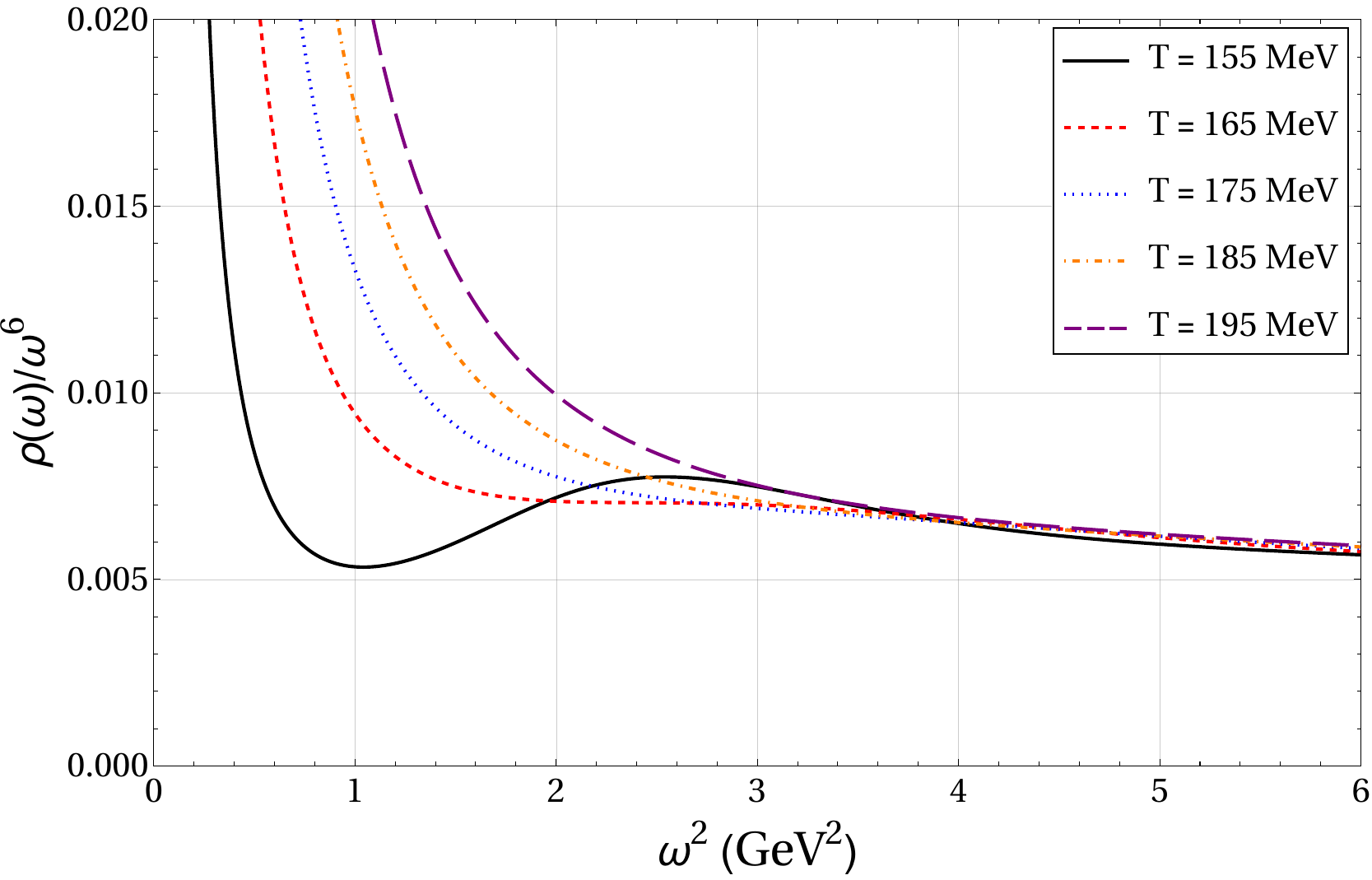}
    \caption{\small Spectral function for various values of temperature in the EPD geometry with $c=0.55~\text{GeV}$ and $a=0.395~\text{GeV}^2$.}
    \label{fig:EPDSpecA}
\end{figure}

\begin{figure}[h!]
    \centering
    \includegraphics[width=0.75\textwidth]{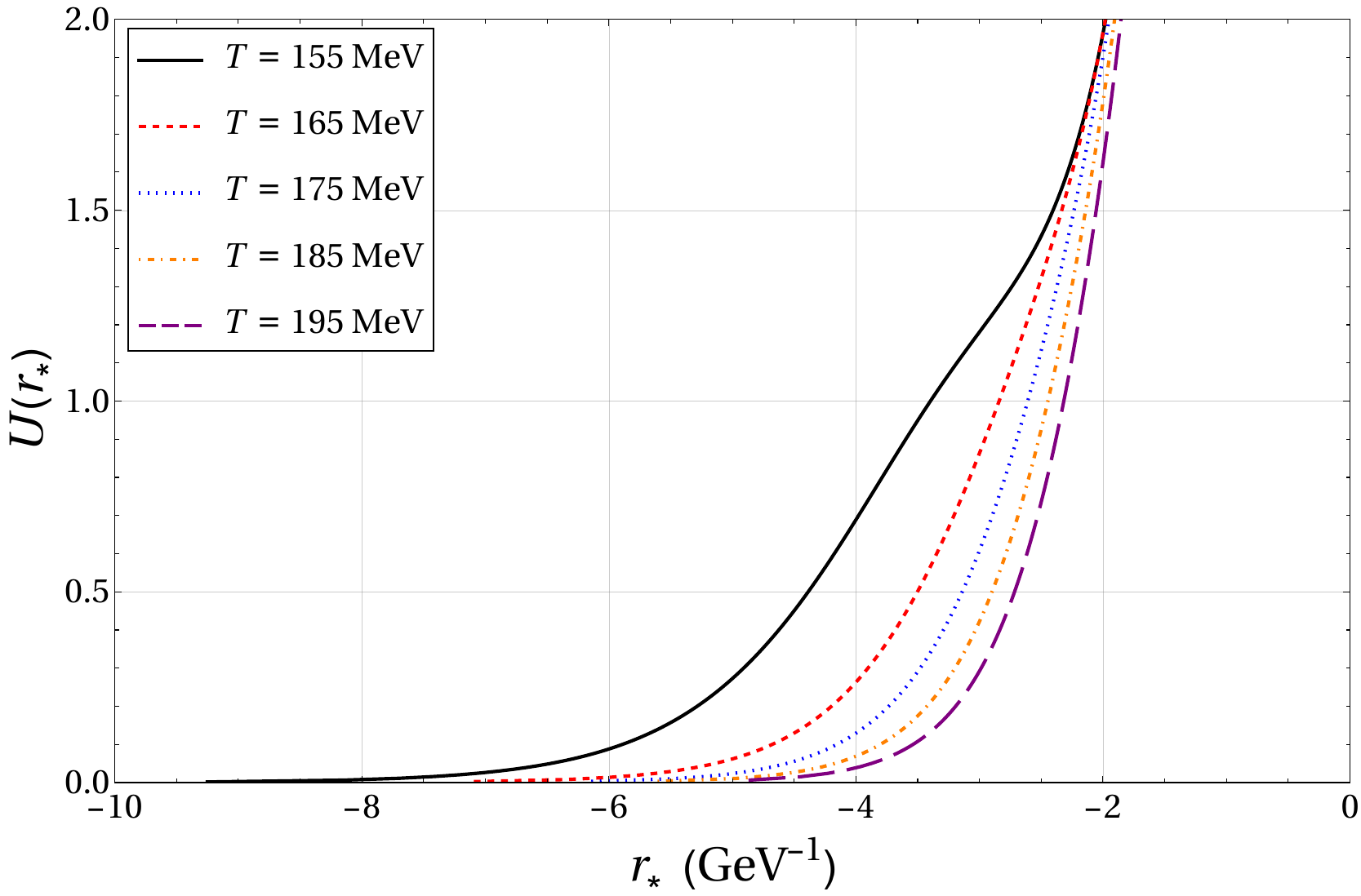}
    \caption{\small Effective potential for various values of temperature in the EPD geometry with $c=0.55~\text{GeV}$ and $a=0.395~\text{GeV}^2$.}
    \label{fig:EPDUA}
\end{figure}

The calculated spectral functions are shown in Fig.~\ref{fig:EPDSpecA} and the associated effective potentials in Fig.~\ref{fig:EPDUA} for a range of temperatures. The latter are expressed in terms of the approximate EPD tortoise coordinate $r_*^{\rm EPD}$ defined above.

In the fully backreacted EPD background, the spectral functions displayed in Fig.~\ref{fig:EPDSpecA} already reflect an advanced stage of in-medium dissociation over the accessible temperature range. Rather than exhibiting a sequence of sharp, well-separated resonant peaks, the spectra are dominated by a comparatively smooth and broadened structure that is gradually flattened as $T$ increases. Thermal widening is still visible, but it is less striking than in the non-backreacting case: since the black-hole branch only exists above a relatively large minimal temperature (fixed by our thermodynamic calibration), the system is effectively probed starting from a regime in which many hybrid-meson excitations are expected to have already melted into the hot plasma. Raising $T$ further produces a modest additional broadening and suppression of the remaining spectral enhancement, instead of a dramatic peak-by-peak disappearance.

A complementary interpretation follows from the associated effective potentials shown in Fig.~\ref{fig:EPDUA}. In the backreacted geometry, the Schr\"odinger potential in the tortoise coordinate does not develop the pronounced confining well that would support long-lived quasi-bound states; already at the lowest available temperature, the potential is comparatively shallow and quickly rises only as one approaches the boundary. As the temperature is increased, the deformation of this profile further weakens any would-be trapping region, reducing the likelihood of narrow resonances and favoring short-lived, continuum-like excitations. This comparatively mild evolution of the potential across the available temperature window naturally explains why the spectral functions in Fig.~\ref{fig:EPDSpecA} broaden less prominently: once a black-hole solution exists, the background already corresponds to a deconfined-like regime, so increasing $T$ mainly refines an already-melted spectrum rather than triggering the onset of melting.

\section{On a finite source for the Proca field} \label{secSource}

A natural question is whether the temporal component of the Proca field, $H_t(z)$, can be treated in analogy with the Maxwell field of Einstein--Maxwell-dilaton models at finite density, as in \cite{Toniato:2025gts}. In the Maxwell case, the boundary value of $A_t$ is identified with the chemical potential because the bulk gauge field is dual to a conserved $U(1)$ current on the boundary. In the present Einstein--Proca-dilaton model, however, the bulk field $H_M$ is massive and is dual to the hybrid operator
\begin{equation}
J_\mu^a = \bar q\,T^a\,G_{\mu\nu}\gamma^\nu q\,,
\end{equation}
so that its boundary datum is not associated with a conserved charge. Therefore, a nonzero UV value of $H_t$ should not be interpreted as a chemical potential. Rather, it corresponds to turning on an external source for the operator $J_t$, i.e.
\begin{equation}
\delta S_{\mathrm{boundary}}=\int d^4x\, h_s\,J_t(x)\,.
\end{equation}
where $h_s$ is the source coupled to the time component of the hybrid operator. This is qualitatively different from the grand-canonical deformation $-\mu Q$ associated with a conserved charge. This distinction is essential: while a Maxwell field provides the standard grand-canonical control parameter for finite-density physics, a Proca field instead defines a source-deformed theory \cite{Korovin:2013bua,Taylor:2015glc}.

The UV behavior of $H_t$ makes this interpretation even more transparent. Since the Proca mass is fixed by the AdS/CFT relation to $m_5^2L^2=8$, corresponding to an operator of dimension $\Delta=5$, the asymptotic expansion of $H_t$ may be obtained directly from the near-boundary form of its equation of motion (\ref{proc}). To do so, we write the relevant quantities at leading order. As $z\to0$, we get
\begin{equation}
A(z)=\mathcal{O}(z^2)\,,\qquad g(z)=1+\mathcal{O}(z^4)\,,\qquad f(z)=1+\mathcal{O}(z^2) \,,
\end{equation}

Hence, the only singular terms that survive, at leading order, near the boundary are
\begin{equation} \label{leading}
H_t''-\frac{1}{z}H_t'-\frac{m_5^2L^2}{z^2}H_t= 0\,.
\end{equation}

Inserting the Frobenius Ans\"atze $H_t(z)\sim z^\alpha $ into the leading equation (\ref{leading}) gives us
\begin{equation}
\alpha(\alpha-1)z^{\alpha-2}-\alpha z^{\alpha-2}-m_5^2L^2 z^{\alpha-2}=0\, ,
\end{equation}
or, equivalently, the indicial equation
\begin{equation}
\alpha(\alpha-2)-m_5^2L^2=0 \,.
\end{equation}

Using $m_5^2L^2=8$, one finds
\begin{equation}
\alpha^2-2\alpha-8=0
 \Longrightarrow 
\alpha_-=-2\,,\,\,\, \alpha_+=4 \,.
\end{equation}

Therefore, the two independent near-boundary behaviors are $H_t(z)\sim z^{-2}$ and $H_t(z)\sim z^{4}$, so the general UV expansion takes the form
\begin{equation}\label{expform}
H_t(z)=h_s\,z^{-2}+h_v\,z^4+\cdots \, .
\end{equation}
where $h_s$ is the coefficient of the non-normalizable mode and is interpreted as the source for the operator $J_t$, whereas $h_v$ multiplies the normalizable mode and is associated with the corresponding response or expectation value. The dots denote subleading terms fixed recursively by the full equation once the explicit forms of $A(z)$, $g(z)$ and $f(z)$ are inserted; depending on the detailed subleading structure, these corrections may include higher powers of $z$ and, in resonant cases, logarithmic terms. For the present discussion, however, only the leading source and response terms are needed.

This implies that the appropriate boundary datum is the coefficient $h_s=\lim_{z\to 0} z^2 H_t(z)$, not a finite value of $H_t(0)$. In particular, the source has dimension $4-\Delta=-1$, so the deformation is irrelevant from the ultraviolet point of view. This is qualitatively different from the Maxwell case, and it already suggests that a finite Proca source is not the natural variable for finite-density thermodynamics \cite{Korovin:2013bua,Baggio:2011cp,Taylor:2015glc}.

Once the functions $A(z)$ and $f(z)$ are fixed by the reconstruction scheme, shown in Section \ref{modelsec}, the coupled background equations (\ref{proc}) and (\ref{metriceq}) may be written as
\begin{equation}
H_t''-\left(\frac{1}{z}+2c^2 z\right)H_t'
-\frac{m_5^2L^2 e^{-a z^2/4}}{g\,z^2}H_t=0 \,,
\end{equation}
\begin{equation}
g''-\left(\frac{3}{z}+\frac{3a}{4}z\right)g'
-\frac{z^2}{L^2}e^{(3a/8-c^2)z^2}\,(H_t')^2
-\frac{m_5^2}{g}e^{(a/8-c^2)z^2}\,H_t^2=0 \,.
\end{equation}

These equations show that the metric reacts to the Proca field only quadratically, through $(H_t')^2$ and $H_t^2$. For this reason, the most natural small-source expansion is
\begin{equation}
H_t(z)=\varepsilon\,h(z)\,, \qquad
g(z)=g_0(z)+\varepsilon^2 g_2(z)+O(\varepsilon^4)\, .
\end{equation}
with $\varepsilon\propto h_s$. At leading order, one first solves for the zero-source background $g_0(z)$, then determines $h(z)$ from a linear equation on that background, and finally computes the first correction $g_2(z)$ from a linear inhomogeneous equation sourced by $h^2$ and $(h')^2$. This perturbative organization makes clear why, for a sufficiently small source, the geometry remains very close to the $H_t=0$ background even when the Proca profile itself varies appreciably \cite{Cheng:2009df,Baggio:2011cp}.

The numerical problem associated with a finite Proca source is correspondingly difficult. First, because the sourced branch behaves as $H_t\sim h_s z^{-2}$, even small changes in $h_s$ are amplified by the factor $1/z^2$ near the UV cutoff. Second, the equations contain inverse powers of the blackening function $g(z)$, while the horizon condition requires $g(z_h)=0$. As a consequence, direct boundary-value or shooting methods are stiff and highly sensitive to the choice of cutoff and initial data. Third, the irrelevance of the source means that the fully backreacted sourced solution is intrinsically UV delicate, in close analogy with the ultraviolet sensitivities discussed in Einstein--Proca holography more generally \cite{Cheng:2009df,Baggio:2011cp,Chemissany:2014xpa,Taylor:2015glc}. In practice, this makes a brute-force nonlinear construction with finite $h_s$ much less robust than either the source-free problem or a controlled perturbative expansion around $h_s=0$.

These observations suggest a clear strategy. If the main goal is the spectroscopy and thermal behavior of hybrid states, the most robust procedure is to keep the Proca source switched off in the background and study the hybrid channel through fluctuations. If one wishes to investigate finite-density effects, the more physical extension is to add a separate Maxwell sector dual to a conserved charge, while keeping the Proca field as the hybrid degree of freedom. A finite Proca source remains meaningful as an external deformation, and it may be useful for studying linear response, susceptibilities, or source-deformed spectral functions in the exotic channel; however, it should not be regarded as a chemical potential, and its fully backreacted treatment is numerically and conceptually much more delicate than in the Maxwell case \cite{Toniato:2025gts,Korovin:2013bua,Baggio:2011cp,Taylor:2015glc}.

\section{Concluding remarks} \label{conc}
\label{sec5}

{\color{black}{In this work, we have constructed a self-consistent holographic model for exotic $1^{-+}$ hybrid mesons within a fully backreacted EPD setup. Starting from the five-dimensional EPD action, we derived and solved the coupled  equations of motion through a potential-reconstruction approach, ensuring dynamical consistency of the background geometry. At zero temperature, we reduced the Proca dynamics to a Schrödinger-like equation and computed the hybrid-meson mass spectrum both in the pure soft-wall AdS limit and at the fully backreacted geometry. We then extended the analysis to finite temperature, analyzing the black hole thermodynamics, fixing the parameter $a$ through the Hawking--Page transition temperature, and evaluating the corresponding spectral functions in both AdS–Schwarzschild and EPD backgrounds.

Our results demonstrate the nontrivial role of gravitational backreaction on observable quantities in holographic QCD. Although the effective Schrödinger potentials and wavefunctions in the AdS and EPD geometries are qualitatively similar, the inclusion of the massive Proca field in a self-consistent setup modifies the background warp factor, introducing exponential corrections to the bulk mass term. These effects translate into noticeable quantitative shifts in the hybrid meson Regge trajectory. In particular, the fully backreacted EPD model yields a ground-state mass for the $\pi_1(1400)$ hybrid meson resonance closer to experiment and slightly alters the slope and intercept of the Regge trajectory compared to the non-backreacting soft-wall result. Hence, solving the EPD field equations is not merely a formal improvement but leads to concrete changes in phenomenological predictions.
Also, regarding the backreacted EPD model, the mass spectrum of the $\pi_1(1600)$ and $\pi_1(2015)$ hybrid meson resonances is closer to the experimental mass spectrum reported in PDG \cite{ParticleDataGroup:2024cfk}.

At finite temperature, the thermodynamic analysis reveals a first-order Hawking--Page transition between thermal AdS (confined phase) and a large black hole geometry (deconfined phase), with the transition temperature fixed to $T_c \simeq 150\,\text{MeV}$ by comparison with lattice QCD. In the non-backreacting AdS–Schwarzschild case, the spectral functions display the expected thermal melting pattern, comprising a well-defined resonance which progressively broadens and disappears as the temperature increases, reflecting the dissociation of hybrid bound states in a hot medium. The corresponding effective potential becomes increasingly shallow, eventually losing the well structure required to support quasi-bound states.

On the other hand, in the fully backreacted EPD geometry, black hole solutions exist only above a relatively large minimal temperature. Consequently, the system is probed directly in a regime that already resembles a deconfined QGP. The spectral functions in this case exhibit broadened, less sharply defined structures from the outset, and increasing temperature produces a comparatively milder additional suppression. Physically, this indicates that hybrid mesons are strongly affected by the thermal-medium effects once the deconfined phase is reached, and that much of the melting process has effectively occurred by the time a stable black hole branch is thermodynamically favored. The associated effective potential supports this interpretation, as it does not develop a deep confining well in the relevant temperature range. 
Our analysis shows that a dynamically consistent EPD background leads to measurable modifications in both vacuum spectroscopy and in-medium behavior of hybrid mesons. The results emphasize the importance of gravitational backreaction in holographic QCD models and suggest that exotic states, such as $1^{-+}$ hybrid meson resonances, are particularly sensitive probes of the interplay between confinement dynamics and thermal deconfinement. A direct comparison with lattice QCD is presently limited by the absence of corresponding finite-temperature spectral results for the light $J^{PC}=1^{-+}$ hybrid channel. Existing finite-temperature lattice studies concern instead the heavy charmonium $1^{-+}$ state~\cite{Urrea-Nino:2025ijv}.

Future extensions of our presented framework may include finite chemical potential (by addition of a Maxwell field), nonzero spatial momentum (distinguishing longitudinal and transverse modes), and the study of quasinormal modes to further clarify the real-time properties of exotic hadrons in strongly coupled matter. 

\section*{Acknowledgments}
BT thanks the São Paulo Research Foundation (FAPESP) Grants No.  2025/28027-7 and No. 2021/01089-1. HA is supported by FAPESP Grant No. 2025/01268-4 and No. 2021/01089-1. RdR~thanks FAPESP (Grants No. 2021/01089-1, No.  2025/23004-9, and No. 2026/14943-4), and the National Council for Scientific and Technological Development (CNPq) (Grants No. 303742/2023-2 and No. 401567/2023-0), for partial financial support.  The work of DD was supported by KU Leuven IF project~C14/26/108.

\bibliographystyle{JHEP}
\bibliography{mybib}
\end{document}